\documentclass[12pt,a4paper]{article}
\usepackage{amssymb,amsmath,graphicx,subcaption,hyperref,cite}
\usepackage{epstopdf}
\usepackage[utf8]{inputenc}
\usepackage[english]{babel}
\usepackage{soul}
\usepackage{color}
\usepackage{amsmath}
\begin{document}

\begin{center}
 {\Large\bf Phase transition in Kermack-McKendrick Model of Epidemic: Effects of Additional Nonlinearity and Introduction of Medicated Immunity}
\end{center}
\vskip 1 cm
\begin{center} %
 Agniva Datta$^1$ and Muktish Acharyya$^2$
  
 \textit{Department of Physics, Presidency University,} 
  
 \textit{86/1 College Street, Kolkata-700073, India} 
 \vskip 0.2 cm
 \textit{Email$^1$:agnivadatta98@gmail.com}
  
 \textit{Email$^2$:muktish.physics@presiuniv.ac.in}
\end{center}
\vspace {1.0 cm}
\vskip 0.5 cm
\noindent {\large\bf Abstract:}
Mathematical modelling of the spread of epidemics has been an interesting challenge in the field of epidemiology. The SIR Model proposed by Kermack and McKendrick in 1927 is a prototypical model of epidemiology. However, it has its limitations. In this paper, we show two independent ways of generalizing this model, the first one if the vaccine isn't discovered or ready to use and the next one, if the vaccine is discovered and ready to use. In the first part, we have pointed out a major over-simplification, i.e., assumption of variation of the time derivatives of the variables with the linear or quadratic powers of the individual variables and introduce two new parameters to incorporate further nonlinearity in the number of infected people in the model. As a result of this, we show how this additional nonlinearity, in the newly introduced parameters, can bring a significant shift in the peak time of infection, i.e., the time at which the infected population reaches maximum. We show that in special cases, even we can get a transition from epidemic to a non-epidemic stage of a particular infectious disease. We further study one such special case and treat it as a problem of phase transition. Then, we investigate all the necessary parameters of this phase transition, like the order parameter and critical exponent. We observe that $O_p \sim (q_c-q)^{\beta}$. {\it As far as we know the phase transition and its quantification in terms of the scaling behaviour is not yet know in the context of pandemic}. In the second part, we incorporate in the model, a consideration of artificial herd immunity and show how we can decrease the peak time of infection with a subsequent decrease in the maximum number of infected people. Finally, we estimate a critical value of the rate of vaccination by a statistical method. Thereby, we propose a possible way of eradicating the epidemic in a short time by effectively providing the vaccine to a population.


\newpage

\section{Introduction}
The spreading of an epidemic is an interesting subject in modern research of 
sociophysics\cite{psbkc}.
The conventional definition of an epidemic is given as a fast spread of disease to a large number of species in a particular population within a little amount of time. The first well-documented epidemic report is of the Justinian Plague, back in 541 A.D. After that, several epidemics have hit mankind through the ages, resulting in a significant fluctuation in the global population, the most recent being the ongoing SARS-COVID-19 epidemic. From as early as the seventeenth century, several sincere attempts have been made to understand the spread of these epidemics in the light of mathematical equations. But the most successful model was proposed by Anderson Gray McKendrick and William Ogilvy Kermack in the year 1927\cite{kermack}. Popularly known as the SIR Model, it successfully explained almost every epidemic in the literature to a considerable extent and is still equally relevant. Of them, Influenza epidemic data for a boys' boarding school as reported in the British medical journal, The Lancet, 4th March 1978 and Bombay Plague of 1905-06 deserve special mention. Most of the epidemic models that followed after this, were based on different modifications of the SIR Model.

Let us briefly review the recent studies in the context of COVID-19. The novel coronavirus (COVID-19), which causes an
acute respiratory disease in human (may be fatal) has spread
to many countries all over the world and has already been declared
as a pandemic by the World Health Organisation\cite{covid1,covid2}. A considerable number of analyses of the available data
of the number of cases and deaths have been attempted
recently, and a few data-driven models have also been
proposed\cite{gaeta,chen,kastner}. The timing of social policies restricting the
social movements is suggested in a recent study\cite{liu}. The transmission of
the disease and spreading by social mixing are studied by computer 
simulation\cite{fang}. A novel fractional time delay dynamic system (FTDD) 
was proposed\cite{cheng}to describe the local outbreak of COVID-19. The evaluation
of the outbreak in Wuhan, China has been analysed\cite{zhau}.
A recent study aimed to establish an early screening model to distinguish COVID-19 pneumonia from Influenza-A viral pneumonia and healthy cases with pulmonary CT images using deep learning techniques\cite{xu}.

The pedagogical studies are also in progress. The scaling features are predicted
\cite{li} in the spreading of COVID-19. The sub-exponential growth of spreading
in Wuhan was reported\cite{maier}. The latent variables, in the auto-encoder and clustering algorithms, are used \cite{hu}to group the provinces for investigating the transmission structure and forecasted curves of cumulative confirmed cases of Covid-19 across China from Jan 20, 2020, to April 20, 2020. The visual data analysis and simulation are done\cite{shi} for prediction of the spread of COVID-19. Epidemic
analysis by dynamic modelling\cite{peng} as well as trend and forecasting of spreading of COVID-19 in China\cite{feng} were also studied. The fractal nature of
the kinetics of COVID-19 was proposed by Ziff\cite{ziff}. The space-time dependence of spreading was also studied\cite{biswas}. The prediction using the SIR model on the Euclidean
network was studied recently\cite{khaleque}.

In this paper, we have modified the Kermack-McKendrick model in the
next section (section-2). The results of our observation are given in section-3 and the paper
ends with concluding remarks in section-4.

\subsection{The SIR Model}
\begin{figure}[h]
 \centering
 \includegraphics[width=0.88\columnwidth]{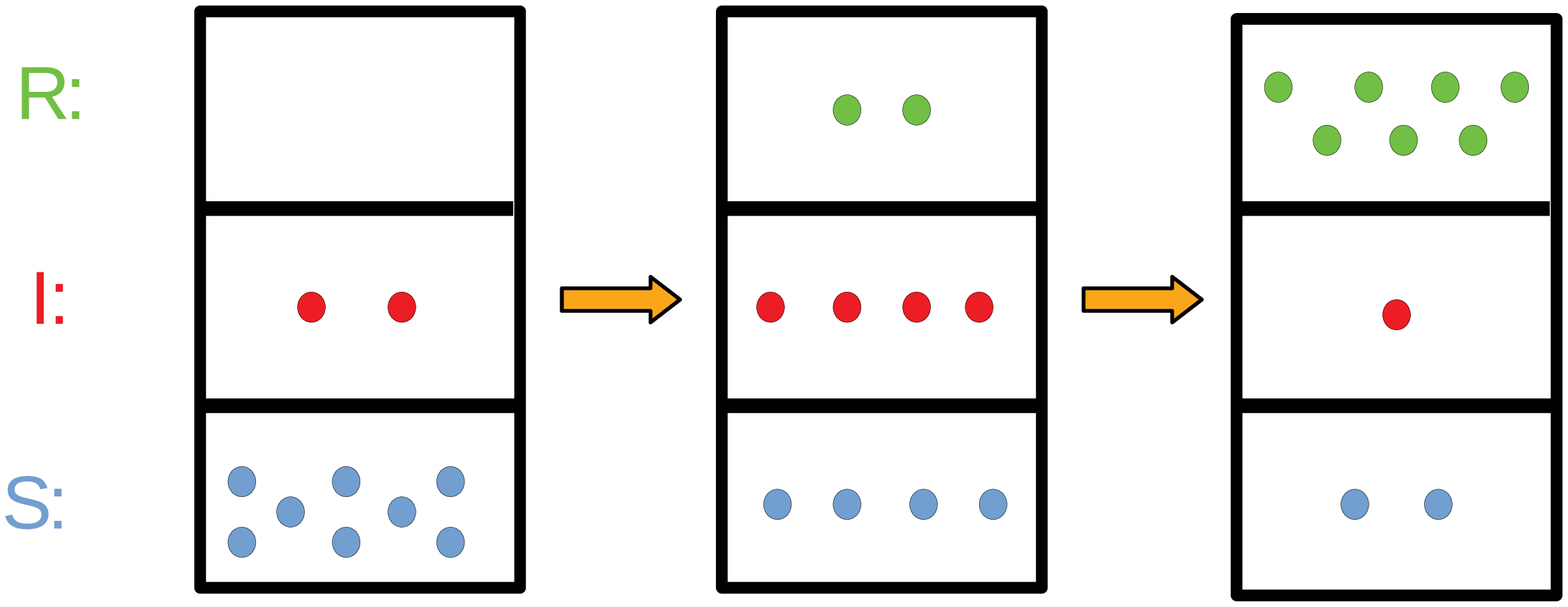}
 \caption{(Color online) Schematic Diagram of the SIR Model showing time evolution of the susceptible(S), infected(I) and removed(R) population.}
 \label{fig:img1}
\end{figure}

\begin{figure}[h]
 \centering
 \includegraphics[width=0.88\columnwidth]{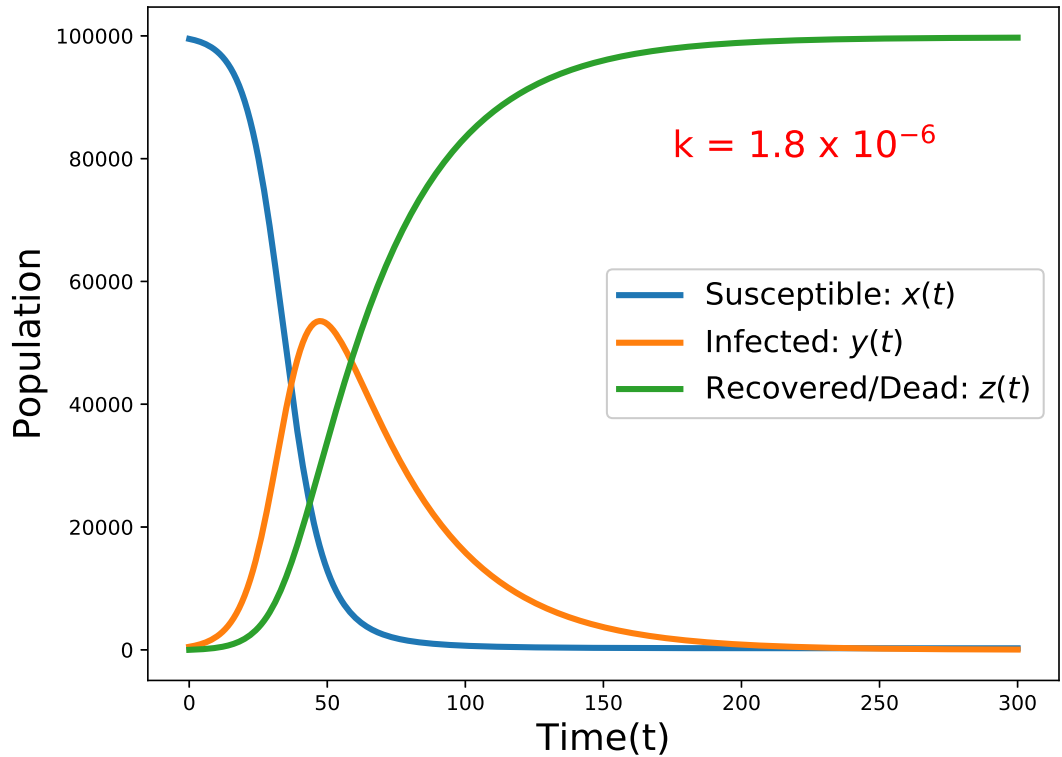}
 \caption{(Color online) The susceptible, infected and removed population are plotted as a function of time for a typical value of $R_{0} > 1$. Parameters are chosen to be $l = 0.03$ and $k = 1.8 \times 10^{-6}$. The susceptible and removed population respectively decreases and increases continuously and saturate. However, the infected population first increases, reaches a peak, which is essentially the $t_{peak}$ and then decreases after the population develops natural herd immunity. This plot is totally at par with the schematic diagram shown in FIG.~\ref{fig:img1}.}
 \label{fig:sir}
\end{figure}

The Kermack-McKendrick Model (1927), also known as the SIR (Susceptible-Infected-Removed) Model\cite{kermack,strogatz} is a compartmental epidemiological model containing three components: the susceptible population, the infected population and the removed (recovered and dead) population. It takes into consideration, three basic assumptions:
(a) the slower changes in population due to births, emigration, deaths by other causes are ignored; (b) the total population remains constant in size (dead people also counted), i.e, the system is conservative; (c) a person once recovered is no more susceptible. Let us consider a toy model to explain the phenomenon. Consider a house consisting of three floors, where a number of people susceptible to infection stay on the ground floor, the ones having infection stay in the middle floor and the recovered and dead people are kept on the top floor. Nobody comes in, nobody goes out but the people in the house can interact freely with each other. We start with a condition, where there are ten people in the house and one of them is infected. As a result of the interactions, as time evolves, more and more susceptible people will get infected and the population on the ground floor will decrease with a subsequent increase in the population of the middle and top floor, as a number of infected people will recover or die and hence, shifted to the top floor. But as time further evolves, there will no longer be enough susceptible people left to acquire infection and as a result, the infected population will decrease. This is known as herd immunity. However, the population of the top floor still increases as it doesn't depend on the number of susceptible people. After a certain time, the infected population will vanish marking the end of the epidemic. A schematic diagram is shown in FIG.~\ref{fig:img1} which explains the compartmental illustration of the model. Mathematically, this model\cite{kermack} is represented as a set of three-variable coupled nonlinear deterministic differential equations given as:

\begin{align}
\frac{dx}{dt} &= -kxy      \nonumber \\
\nonumber \\
\frac{dy}{dt} &= kxy - ly  \label{eqn:SIR} \\ 
\nonumber \\
\frac{dz}{dt} &= ly        \nonumber
\end{align}

where, $x$ = the number of susceptible people, $y$ = the number of infected people, $z$ = number of removed (recovered and dead) people, $k$ and $l$ are the two parameters which are the rate of infection and rate of removal respectively.  The set of coupled differential equations can be solved numerically using the fourth-order Runge-Kutta method. We have plotted the time evolution of $x$,$y$ and $z$ in FIG.~\ref{fig:sir}, assuming $N = 100000$ and $y(t=0) = 500$. 

 A number of suitable substitutions and non-dimensionalization enable us to write Eqn.~\ref{eqn:SIR} in a reduced first-order form given by,
 
 \begin{equation}
     \frac{du}{d\tau} = a - bu - e^{-u}
     \label{eqn:reduced}
 \end{equation}

where, $N = x + y + z =$ total population, $x_{0} = x(t=0) =$ initial susceptible population, $a = \frac{N}{x_{0}}, R_{0} = \frac{kx_{0}}{l}, u = \frac{kz}{l}$ and $\tau = kx_{0}t$.

This $R_{0}$ is a very crucial parameter known as the basic reproduction rate of infection which determines whether an infectious disease will be an epidemic or not.  The following figure, FIG.~\ref{fig:r0} show the time evolution of $\dot{u}$ for two values of $R_{0}$. The following simulation shows the time evolution of $x$, $y$ and $z$ with a change in $k$. Simulation 1: \textcolor{blue}{\url{https://youtu.be/VnajoGwS-4k}}. It is seen from the simulation that the $t_{peak}$, i.e., the time at which the infected population($y$) reaches the maximum, decreases with an increase in the rate of infection($k$), accompanied by an increase in the value of $y$ at $t_{peak}$. So, it is always favourable to delay the $t_{peak}$ so that we can provide sufficient medical infrastructure to the infected people, get enough time for testing and also try to discover the vaccine.

\vspace{0.2cm}

In this paper, we do two modifications to the model independently referring to two situations, one before the discovery of the vaccine and another after the discovery of the vaccine. In the first part, we argue that in addition to the given set of assumptions made in developing the SIR Model, the fact that the rate of change of the susceptible, infected and removed population varies with the linear or quadratic power of $x$, $y$ and $z$, is a bit of over-simplification. Moreover, the parameters $k$ and $l$ take into consideration, both the properties of the agent of infection as well as the host. So, we introduce additional parameters to incorporate nonlinear incidence rates of populations in SIR Model much similar to what was pointed out by Liu et al. \cite{liu1986},\cite{liu1987} and show that we can successfully shift the peak time of infection, and even transform an epidemic disease to a non-epidemic disease by varying the values of these two parameters keeping the values of original parameters of the model, i.e., $k$ and $l$ and hence $R_{0}$, constant. In the situation after the vaccine is discovered and ready to use, we include an additional term that incorporates artificial herd immunity in the model.

\section{Modification of the Model}

\subsection{Introducing further Nonlinearity in Model}
In the papers by Liu et al. \cite{liu1986},\cite{liu1987}, the authors have considered nonlinear incidence rates (in $x$ and $y$) in the SIR Model and pointed out that the consideration can give rise to a variety of dynamical behaviours like saddle-node bifurcations, Hopf bifurcations, etc. which in special cases can give rise to interesting periodic solutions for co-existence of both the host and the disease. However, their effects on the peak time of infection ($t_{peak}$) are not exclusively studied, which bears significant socio-physical as well as epidemiological significance. We introduce two parameters- $p$ and $q$ in the model and name them respectively the Infection Exponent and Removal Exponent, in order to incorporate further nonlinearity in the model. We argue that this not only rectifies the oversimplification but also relaxes the information contained in the two parameters $k$ and $l$ of the original SIR Model. We claim that the parameters $k$ and $l$ are signatures of the characteristics of the agent and the parameters $p$ and $q$ refer to the properties of the host. We modify the equations in Eqn.~\ref{eqn:SIR} as follows:

\begin{align}
\frac{dx}{dt} &= -kxy^{p} \nonumber \\
\nonumber \\ 
\frac{dy}{dt} &= kxy^{p} - ly^{q} \label{eqn:MOD} \\ 
\nonumber \\
\frac{dz}{dt} &= ly^{q} \nonumber
\end{align}

We restrict our model to diseases where everyone is susceptible(like the ongoing COVID-19). It is to be noted that we do not consider nonlinearity in the susceptible population $x$, as $p$ and $q$ are properties of the host (human factors) and hence don't contribute to the susceptibility. Now by varying the values of $p$ and $q$ around 1, we see how the modified model results in the change of $t_{peak}$. We consider two cases, first, by putting $q$ = 1 and varying $p$ and then, by putting $p$ = 1 and varying $q$. \textit{We aim to look for the favourable cases where $t_{peak}$ is shifted to the right so that we get more and more time for the vaccine to be released.}

\subsubsection{Case I: q = 1}

\begin{figure*}[htpb]
 \centering
 (a)\includegraphics[width=0.46\columnwidth]{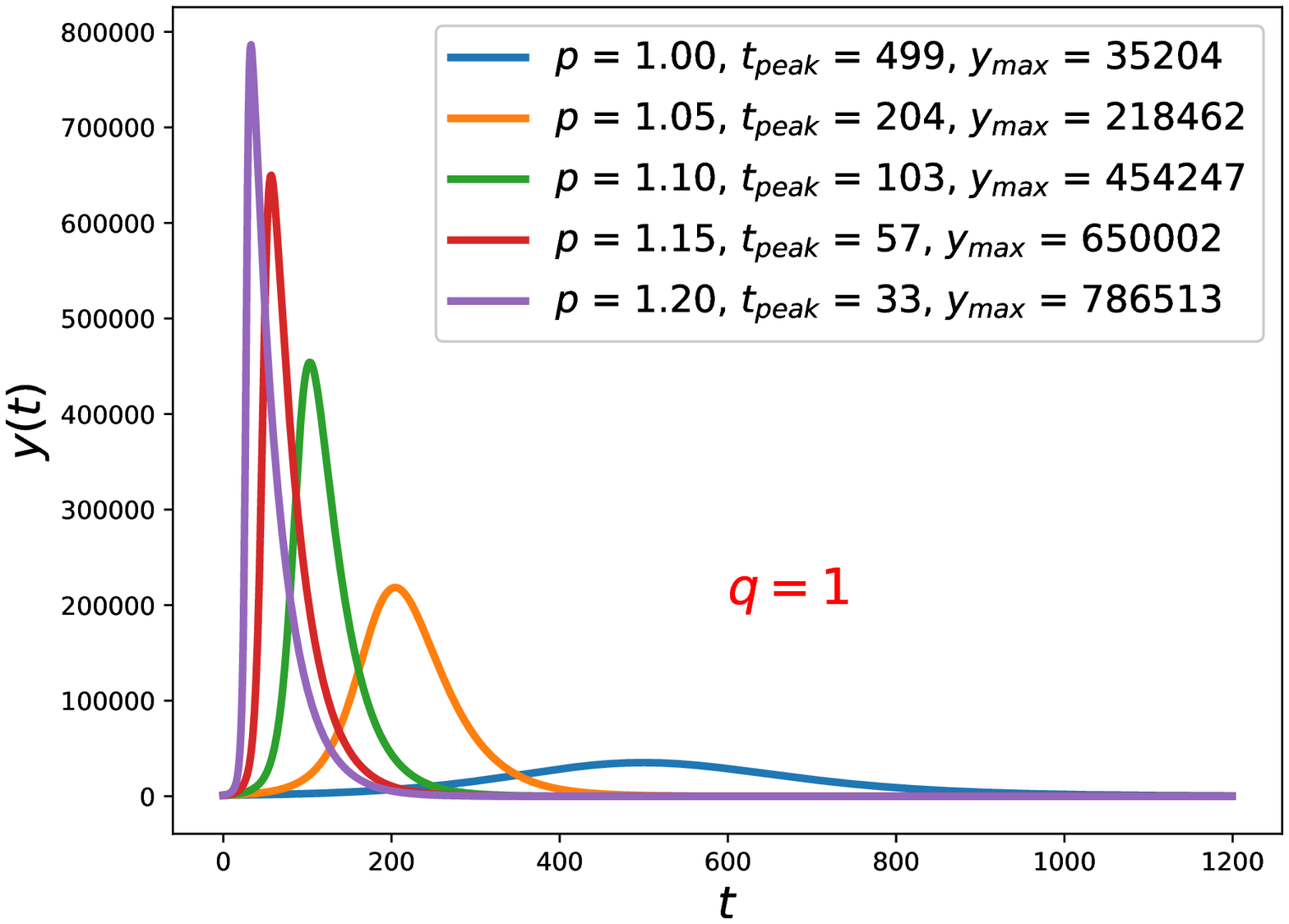}
 (b)\includegraphics[width=0.46\columnwidth]{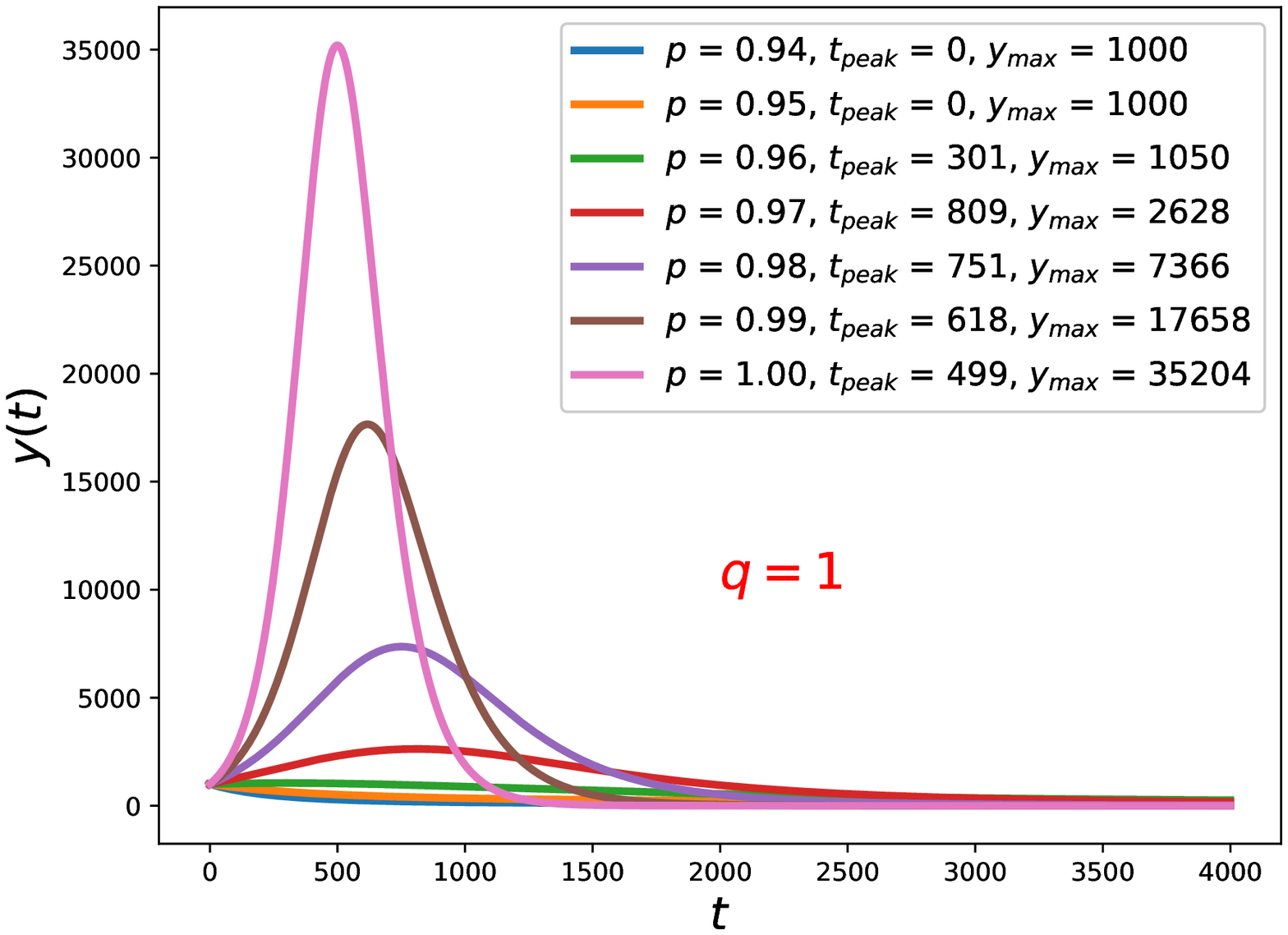}
 \\
 (c)\includegraphics[width=0.46\columnwidth]{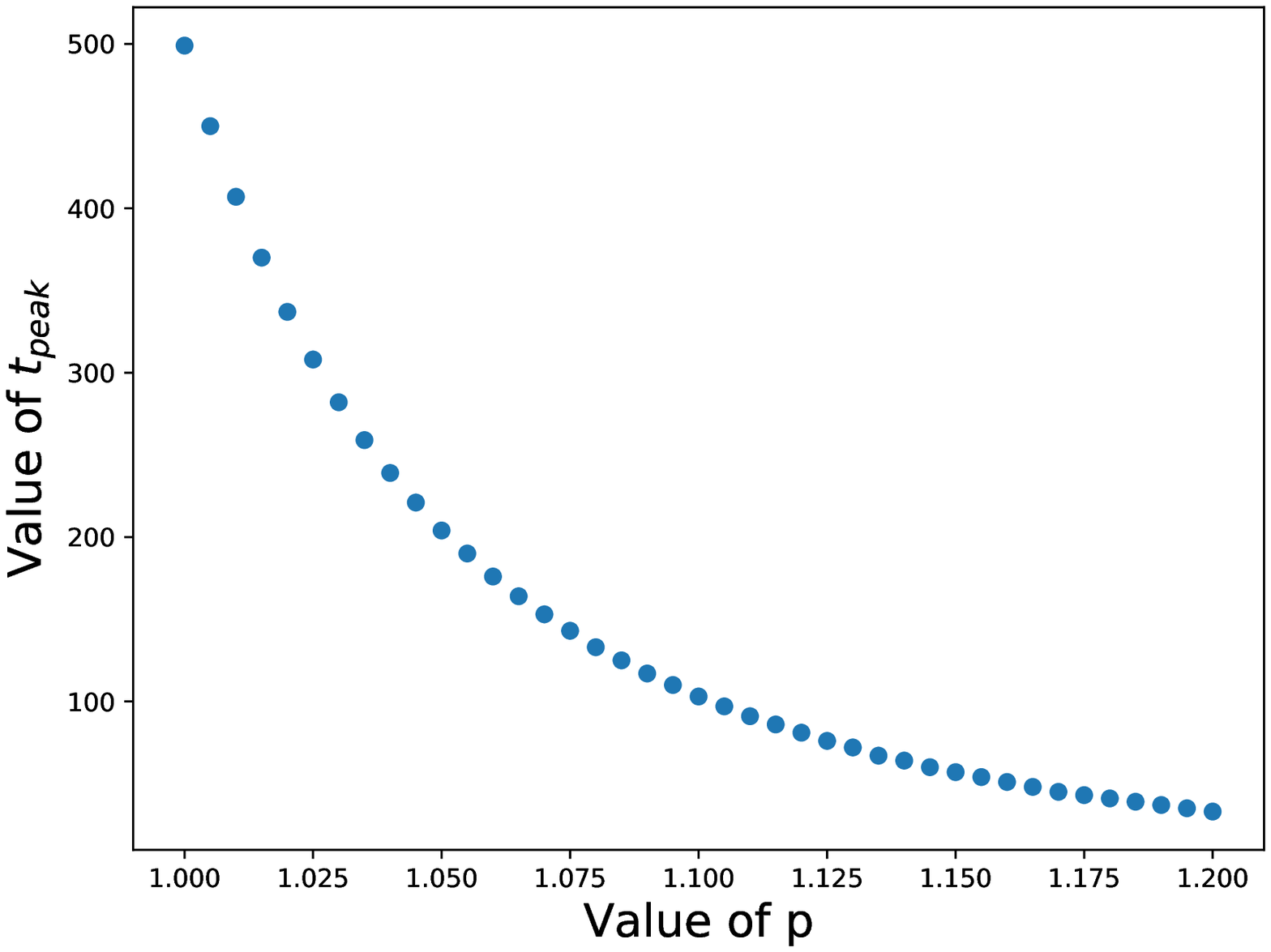}
 (d)\includegraphics[width=0.46\columnwidth]{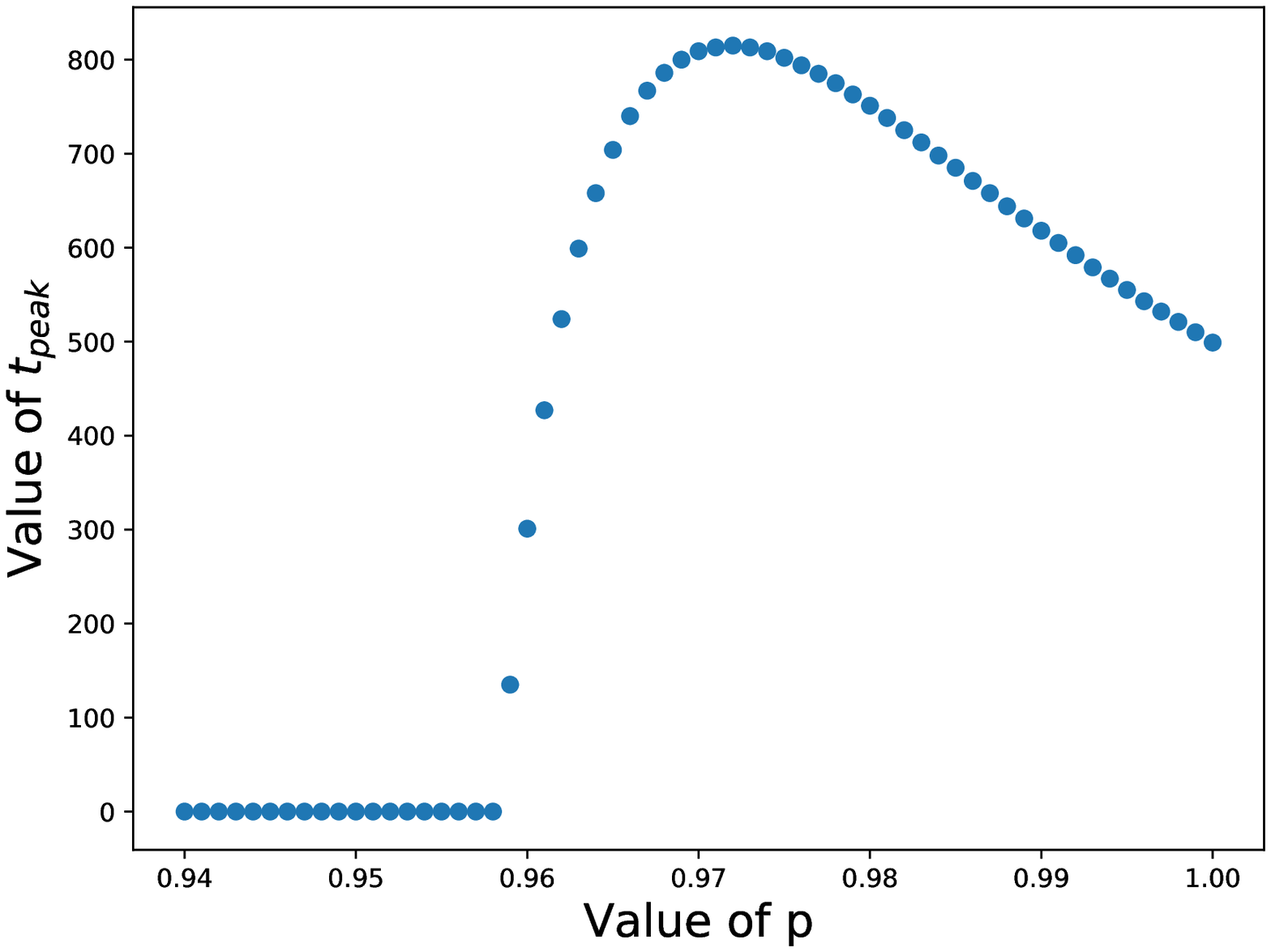}
 \caption{ (Color online) Chosen Parameters: $N = 1000000$, $y(t=0) = 1000$, $k = 4.004 \times 10^{-8}$ (a) We plot $y$ vs. $t$ in super-linear regime of $p$ (b) We plot $y$ vs. $t$ in sub-linear regime of $p$ (c) $t_{peak}$ is plotted as a function of $p$ in super-linear regime (d) $t_{peak}$ is plotted as a function of $p$ in sub-linear regime. }
 \label{fig:q=1}
\end{figure*}

We use fourth order Runge-Kutta method to solve the set of modified equations numerically putting $q$ = 1, for different values of $p$, both in sub-linear ($p <$ 1) and super-linear ($p >$ 1) regime. We plot the time evolution of infected population by changing the parameter $p$ in FIG.~\ref{fig:q=1}(a) and FIG.~\ref{fig:q=1}(b), and compare the values of $t_{peak}$. 

From FIG.~\ref{fig:q=1}(a), i.e., in the super-linear regime, it can be observed that increase in $p$ results in a decline of $t_{peak}$ (shift to the left) with a corresponding increase in the maximum number of infected people (height  of $t_{peak}$), i.e., $y_{max}$. From FIG.~\ref{fig:q=1}(b), i.e., in the sub-linear regime, as $p$ is decreased, $t_{peak}$ is increased (shift to the right) with a decline in $y_{max}$ up to $p$ = 0.97. As $p$ is decreased further, a different trend is observed. A decrease in $t_{peak}$ is observed again but this time, with a corresponding decrease in $y_{max}$ as well. At values of $p$ $<$ 0.96, we see that $t_{peak}$ is zero. A clearer vision of the above can be obtained by looking at the following simulation we have prepared.
Simulation2: \textcolor{blue}{\url{https://youtu.be/ZTHBGCWKOwY}}

\par

For a better understanding of the trend of $t_{peak}$, we plot $t_{peak}$ as a function of the parameter $p$ keeping $q$ = 1 for both the sub-linear and super-linear regime (by taking more data points for precision) in FIG.~\ref{fig:q=1}(c) and FIG.~\ref{fig:q=1}(d) which clearly explains our argument above.

\subsubsection{Case II: p = 1}

\begin{figure*}[htpb]
 \centering
 (a)\includegraphics[width=0.46\columnwidth]{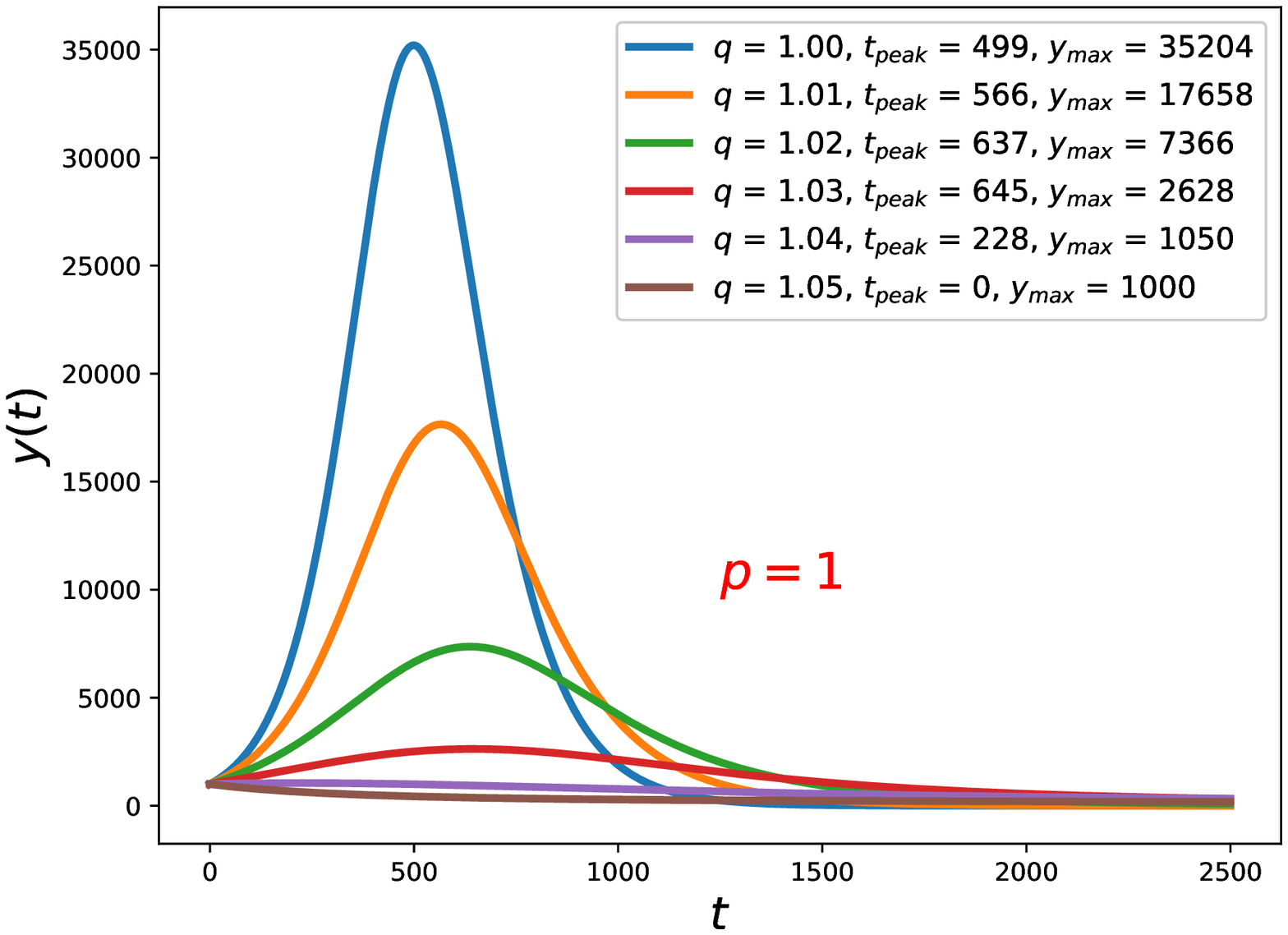}
 (b)\includegraphics[width=0.46\columnwidth]{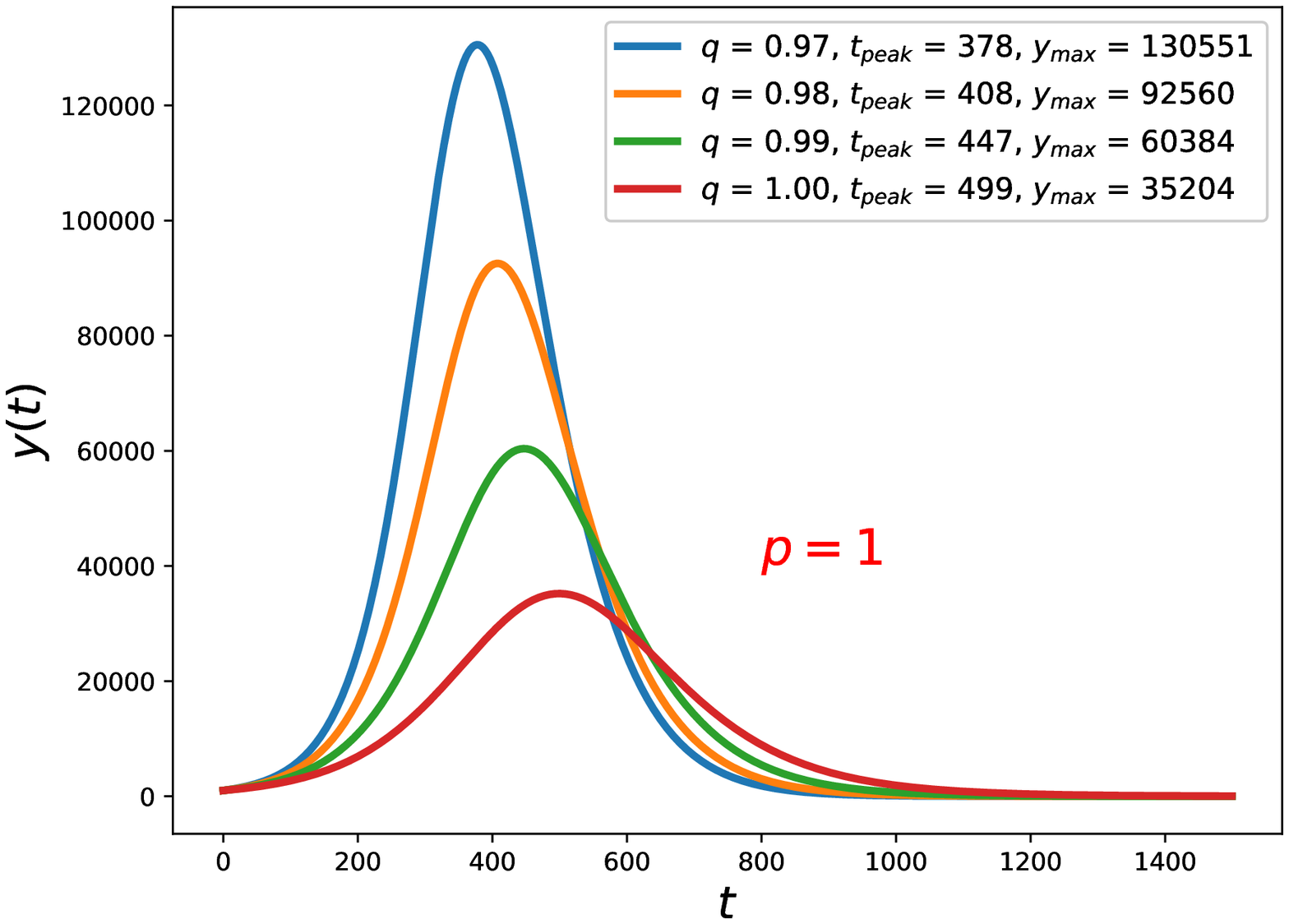}
 \\
 (c)\includegraphics[width=0.46\columnwidth]{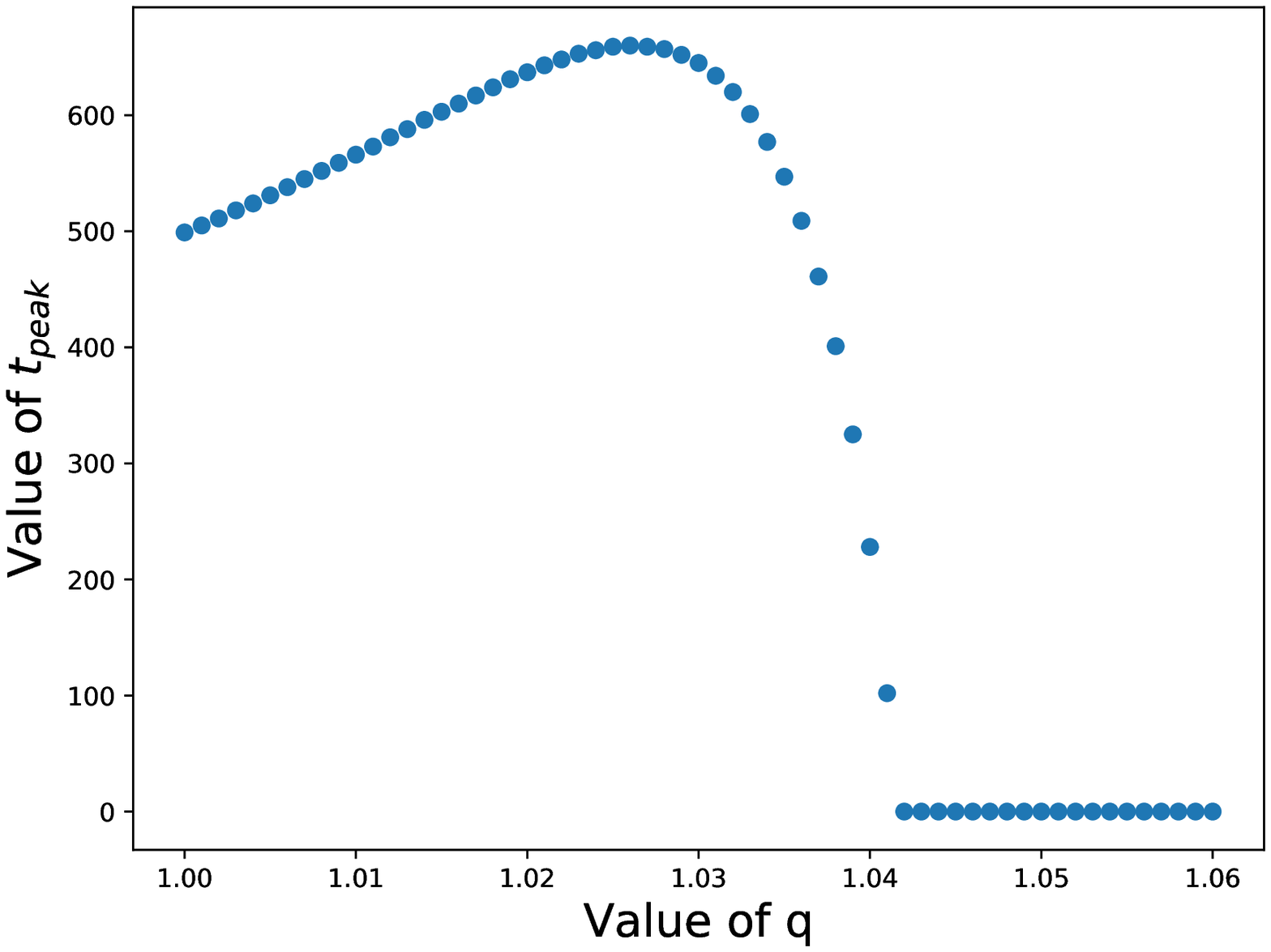}
 (d)\includegraphics[width=0.46\columnwidth]{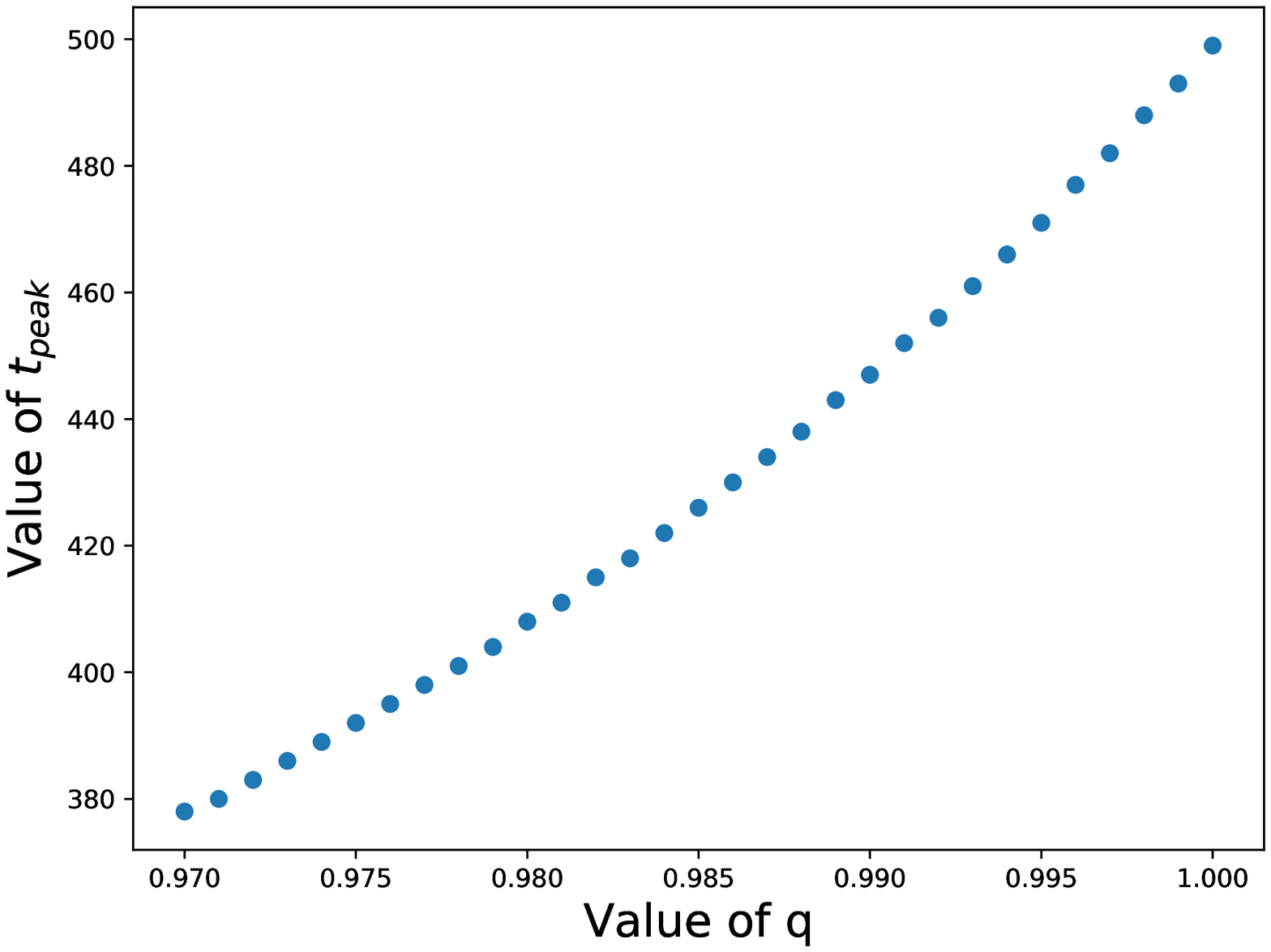}
 \caption{ (Color online) Chosen Parameters: $N = 1000000$, $y(t=0) = 1000$, $k = 4.004 \times 10^{-8}$ (a) We plot $y$ vs. $t$ in super-linear regime of $q$ (b) We plot $y$ vs. $t$ in sub-linear regime of $q$ (c) $t_{peak}$ is plotted as a function of $q$ in super-linear regime (d) $t_{peak}$ is plotted as a function of $q$ in sub-linear regime. } 
 \label{fig:p=1}
\end{figure*}

We take a similar approach to deal with this case following the same initial conditions, now by putting $p$ = 1, for different values of $q$, both in the sub-linear ($q <$ 1) and super-linear ($q >$ 1) regime. The time evolution of infected population by varying $q$ in figures FIG.~\ref{fig:p=1}(a) and FIG.~\ref{fig:p=1}(b) is plotted and the values of $t_{peak}$ are accordingly compared. 

Here, we observe an exactly opposite trend. From FIG.~\ref{fig:p=1}(a), i.e., in the super-linear regime, we observe that increase in $q$ results in increase of $t_{peak}$ (shift to the right) with a corresponding decrease in $y_{max}$ up to $q$ = 1.03. At values of $q >$ 1.03, the trend is similar to that in the sub-linear regime of Case-I. Increase in $q$ results in decrease of $t_{peak}$ with a corresponding decrease in $y_{max}$. At values of $q >$ 1.05, $t_{peak}$ becomes zero. Now in FIG.~\ref{fig:p=1}(b), i.e., in the sub-linear regime, as $q$ is decreased, $t_{peak}$ is decreased (shift to the left) with an increase in $y_{max}$. A similar simulation for this case has also been prepared, which makes the vision clear.
Simulation 3:\textcolor{blue}{ \url{https://youtu.be/81KVIpFEDC0}}

Now similarly, we plot $t_{peak}$ as a function of the parameter $q$ keeping $p$ = 1 for both the sub-linear and super-linear regime (by taking more data points for precision) in FIG.~\ref{fig:p=1}(c) and FIG.~\ref{fig:p=1}(d) and it also successfully explains our argument above. 

\subsection{Introducing Artificial (Medicated) Herd Immunity in the Model}
\begin{figure}[h]
 \centering
 \includegraphics[width=0.88\columnwidth]{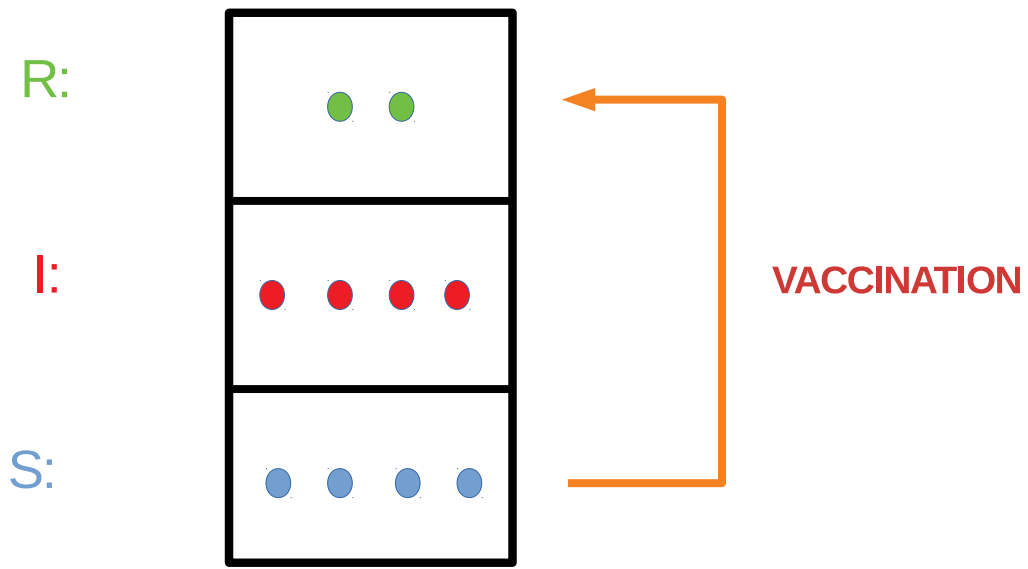}
 \caption{(Color online) Schematic Diagram of incorporating artificial herd immunity in model.}
 \label{fig:img2}
\end{figure}

In the previous subsection, we discussed how the population will evolve if no medication or vaccination is discovered. But if a vaccine is discovered and ready for use, the time evolution of the three populations won't be the same. So, we incorporate artificial (medicated) herd immunity in the SIR model following a particular strategy. In general, vaccination is provided to the population in a constant rate. Shulgin et al. \cite{shulgin} suggested a method of pulsed vaccination which in certain cases require a lower rate of vaccination to eradicate an epidemic. But an ideal condition will be to achieve a state where the susceptibility will fall exponentially with an increase in vaccination rate. Now we know, artificial (medicated) herd immunity means that we essentially transform a fraction of susceptible people to removed people so that they don't acquire infection and are unable to spread it. To incorporate this idea in the model, we introduce a parameter $c$, which corresponds to the rate of vaccination, by which we shift a number of people from the ground floor to the second floor directly without exposing them to the first floor, with reference to FIG.~\ref{fig:img2}. So, we modify Eqn.~\ref{eqn:SIR} as follows:

\begin{align}
\frac{dx}{dt} &= -kxy - cx     \nonumber \\
\nonumber \\
\frac{dy}{dt} &= kxy - ly  \label{eqn:vac} \\ 
\nonumber \\
\frac{dz}{dt} &= ly  + cx     \nonumber
\end{align}

It may be noted here that despite the inclusion of the medication term ($cx$), the system remains conservative ($N=$ fixed). Moreover, as we increase the value of $c$, the second term in the first equation in Eqn.~\ref{eqn:vac} becomes

\begin{equation}
     \frac{dx}{dt} \approx  - cx \Rightarrow x \approx x_{0}e^{-ct}
     \label{eqn:expd}
 \end{equation}

So, in other words, if we can strategically provide the vaccination by considering a medicated herd immunity term proportional to $x$, as given in Eqn.~\ref{eqn:expd}, we can achieve an exponential fall of susceptibility towards the disease. In fact, a logic similar to this model is being followed in a number of countries including India regarding COVID-19, such that the people above the age of 45 and having high blood pressure or Diabetes Mellitus (who are thought to be "more susceptible") are given vaccines earlier.   \par
\begin{figure*}[htpb]
 \centering
 (a)\includegraphics[width=0.75\columnwidth]{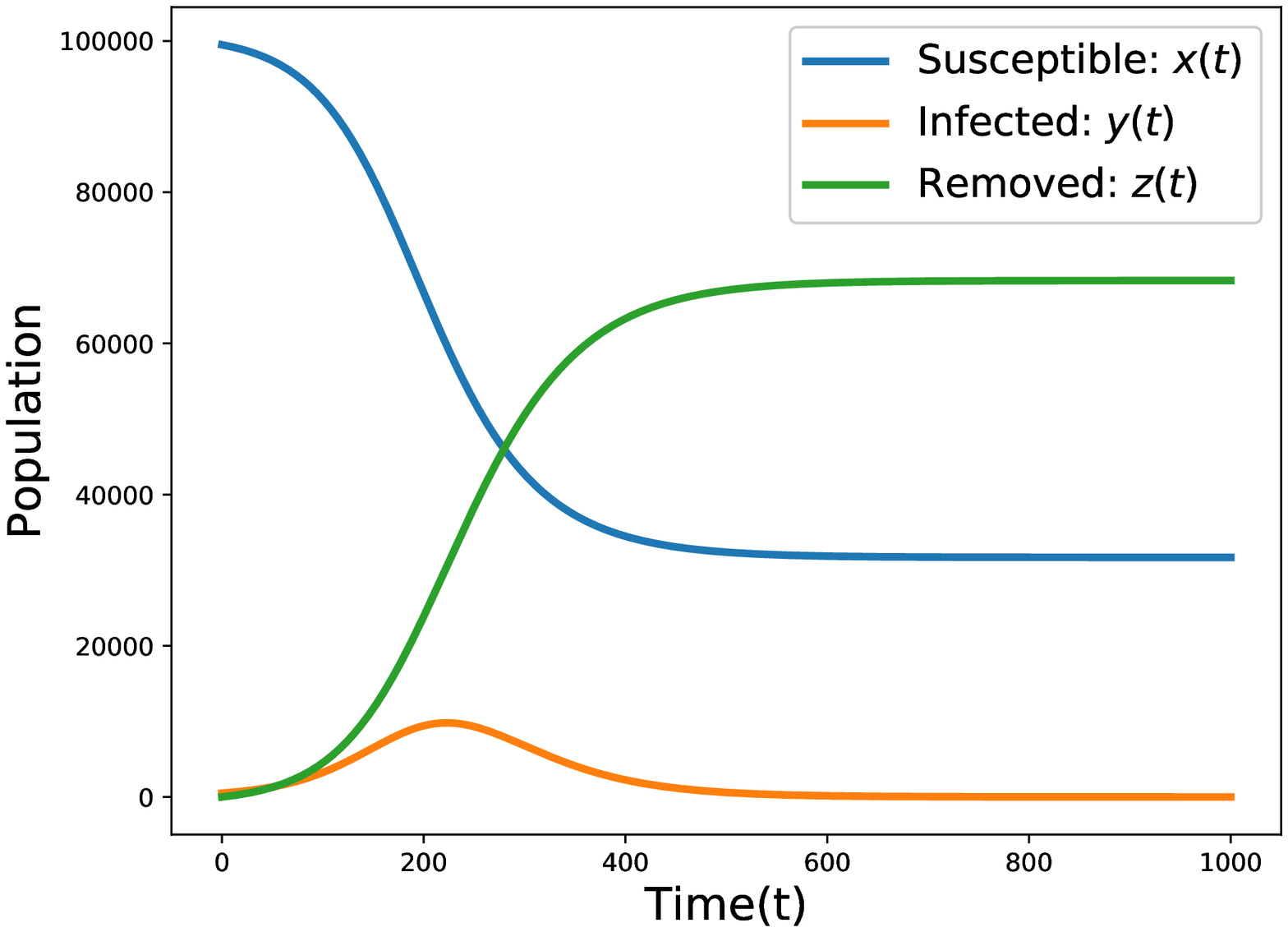}
 (b)\includegraphics[width=0.75\columnwidth]{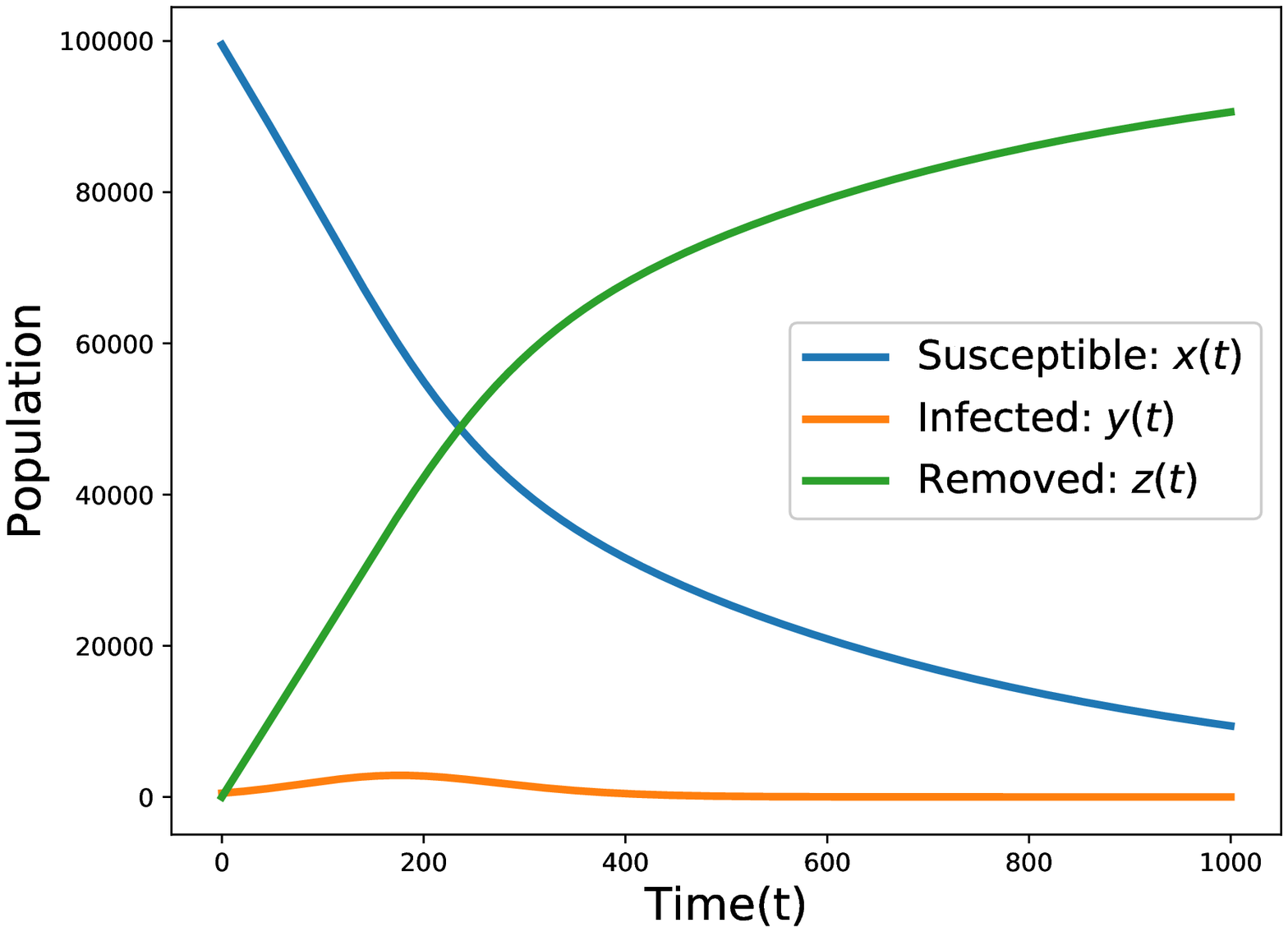}
 \\
 \caption{ (Color online) Parameters chosen: $k = 5.02512563 \times 10^{-7}$, $l$ = 0.03; Plots of $x$, $y$ and $z$ as a function of time (a) in absence of herd immunity ($c$ = 0) (b) in presence of herd immunity ($c = 2 \times 10^{-3}$).} 
 \label{fig:herd}
\end{figure*}

 
We have similarly solved this set of modified equations numerically using fourth-order Runge-Kutta method and plotted the time evolution of $x$,$y$ and $z$ in FIG.~\ref{fig:herd}, assuming $N = 100000$ and $y(t=0) = 500$.It can be clearly seen that for the same value of $k$ and $l$, the nature of time evolution of $x$ and $z$ are different, they tend to decline and grow faster respectively in FIG.~\ref{fig:herd}(b) compared to FIG.~\ref{fig:herd}(a). Although the nature of variation of $y$ remains same, still we see a decrease in both $t_{peak}$ and $y_{max}$ in FIG.~\ref{fig:herd}(b) compared to FIG.~\ref{fig:herd}(a). A simulation is attached in this link which shows the gradual change in the time evolution of $x$, $y$ and $z$ with the increase in $c$, keeping $k$ and $l$ constant. Simulation 4: \textcolor{blue}{\url{https://youtu.be/JkmArmA-pC0}}. A detailed physical interpretation and statistical analysis of this phenomenon are discussed in the next section.

\section{Results and Discussion}
Several significant inferences can be drawn from the modifications of the model in both parts. First, we give a physical interpretation of the plots in the former two sections and then analyze them statistically to understand both the phenomenon in ample depth.

\subsection{Results from Further Nonlinearity in Model}

\subsubsection{Physical Interpretation} \vspace{0.2in}

{\bf{Case-I: q = 1} \par \vspace{0.2in}}

\begin{figure*}[htpb]
 (a)\includegraphics[width=0.46\columnwidth]{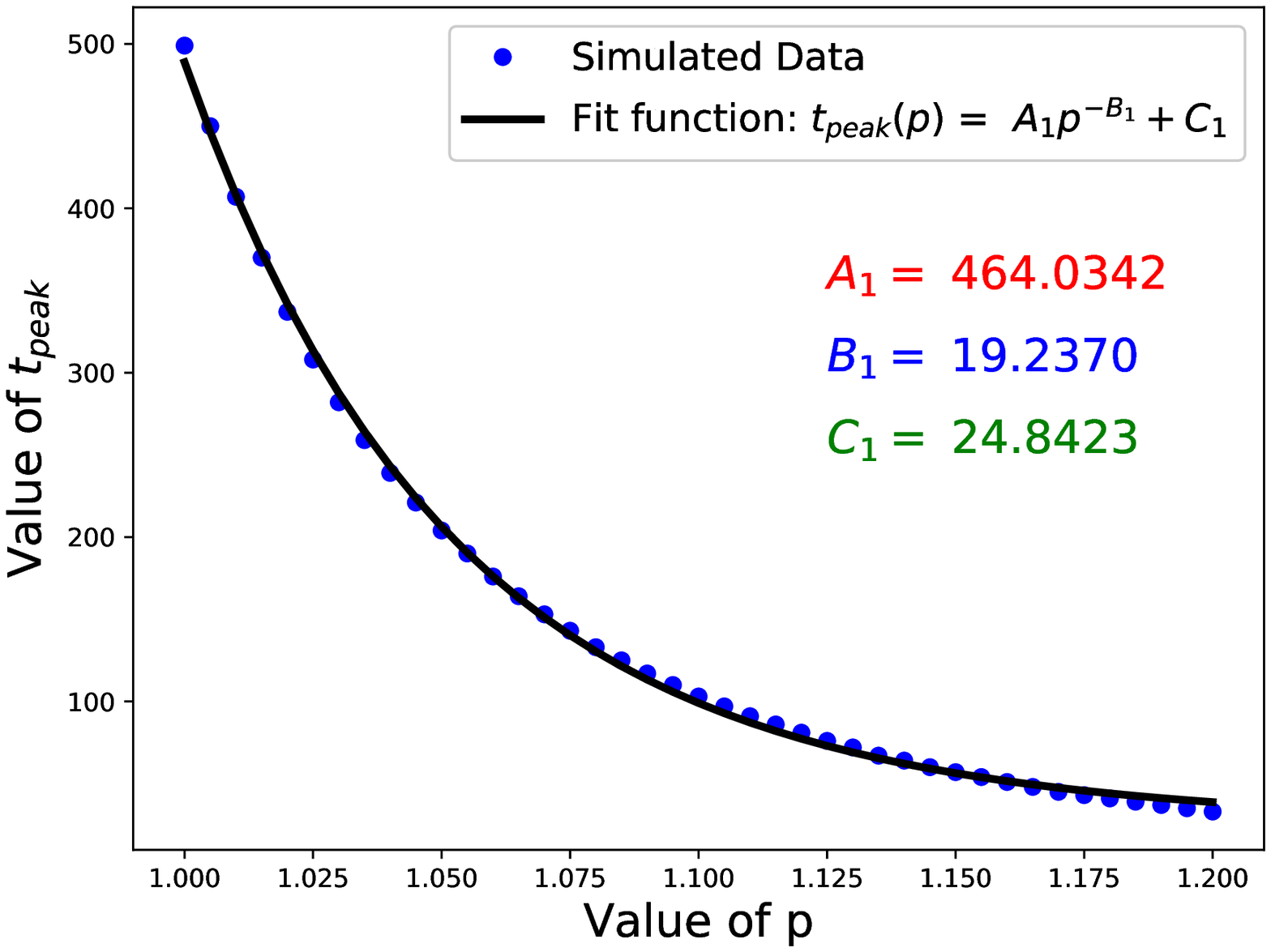}
 (b)\includegraphics[width=0.46\columnwidth]{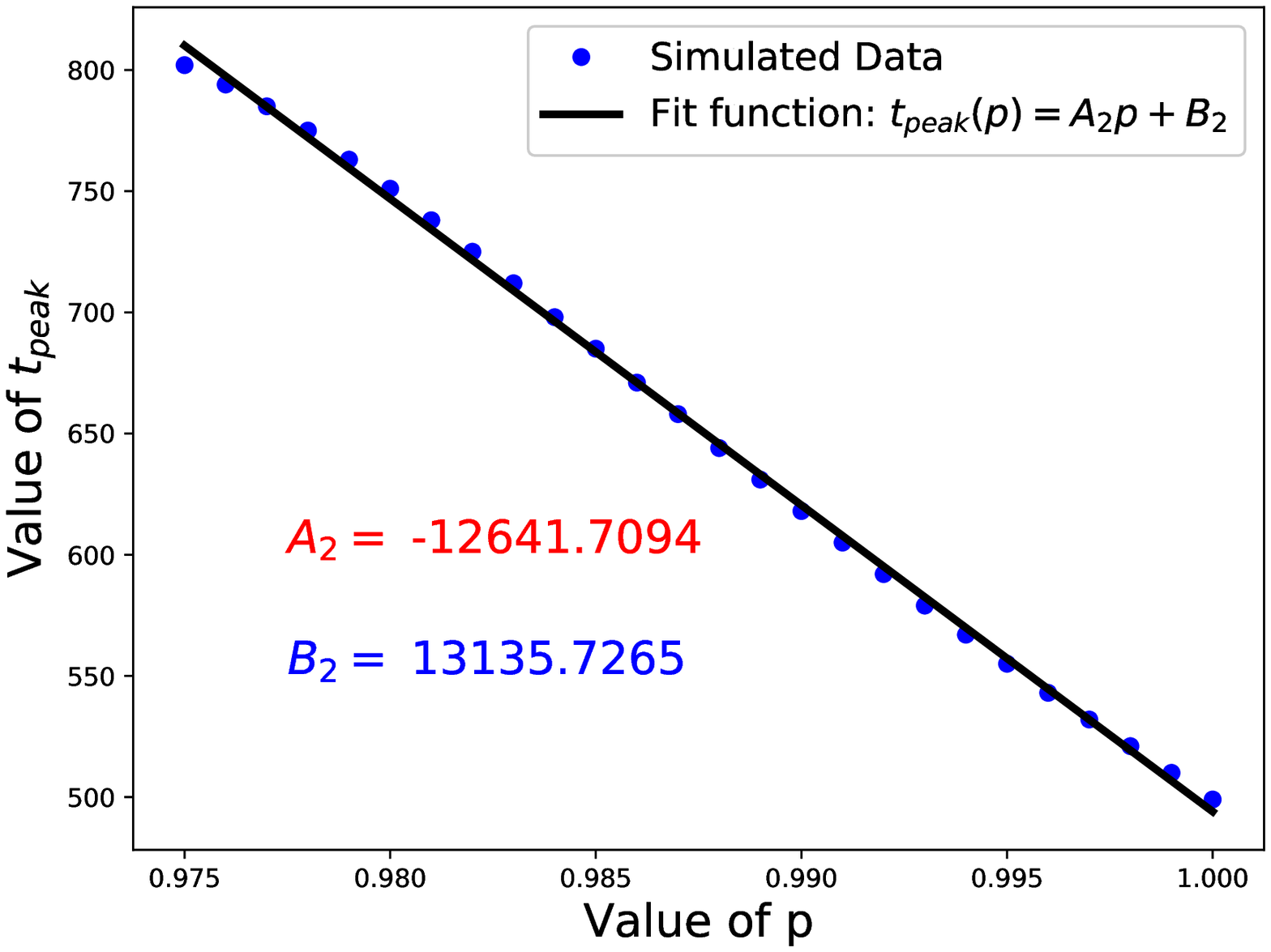} 
 \\
 (c)\includegraphics[width=0.46\columnwidth]{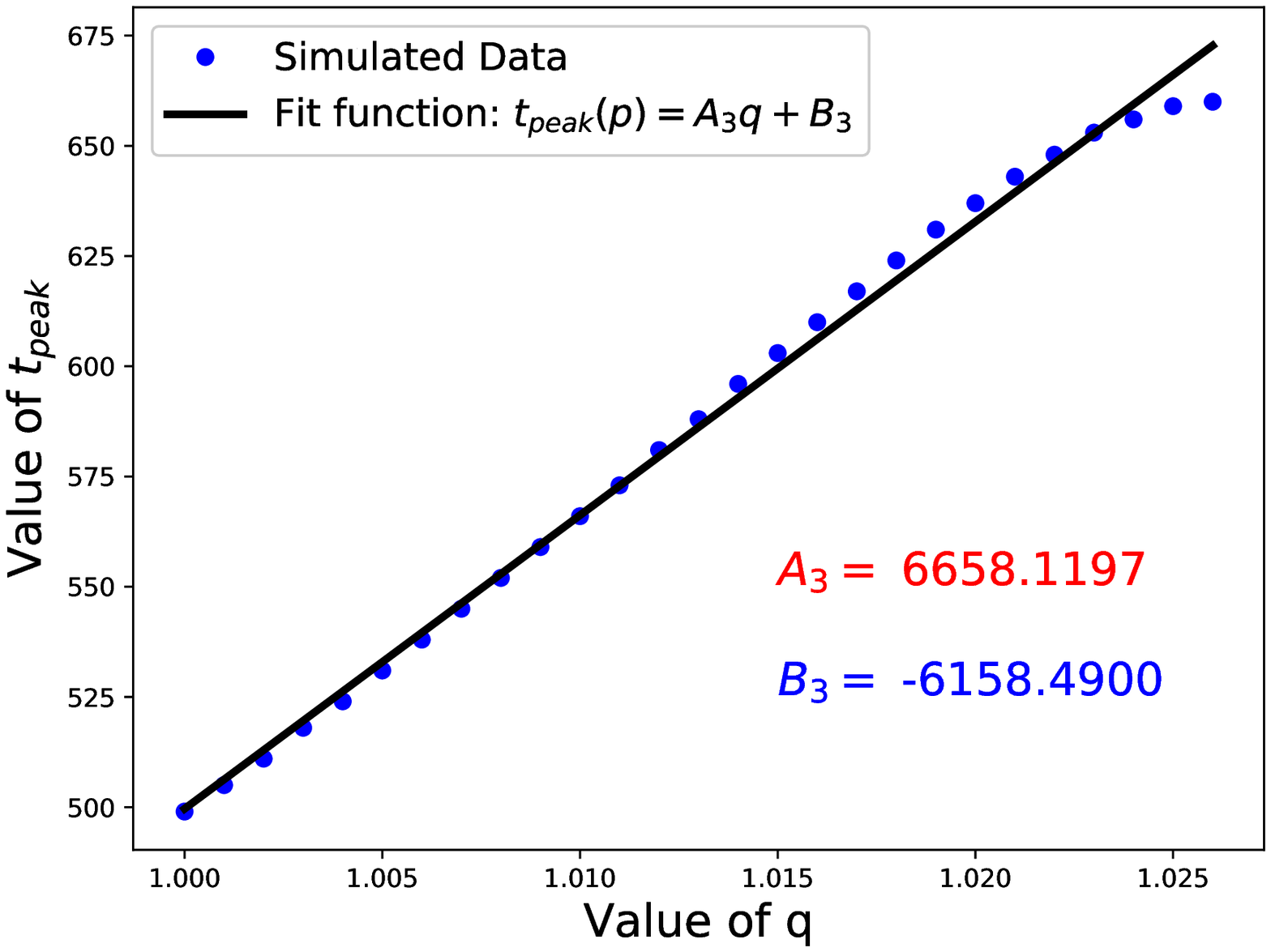}
 (d)\includegraphics[width=0.46\columnwidth]{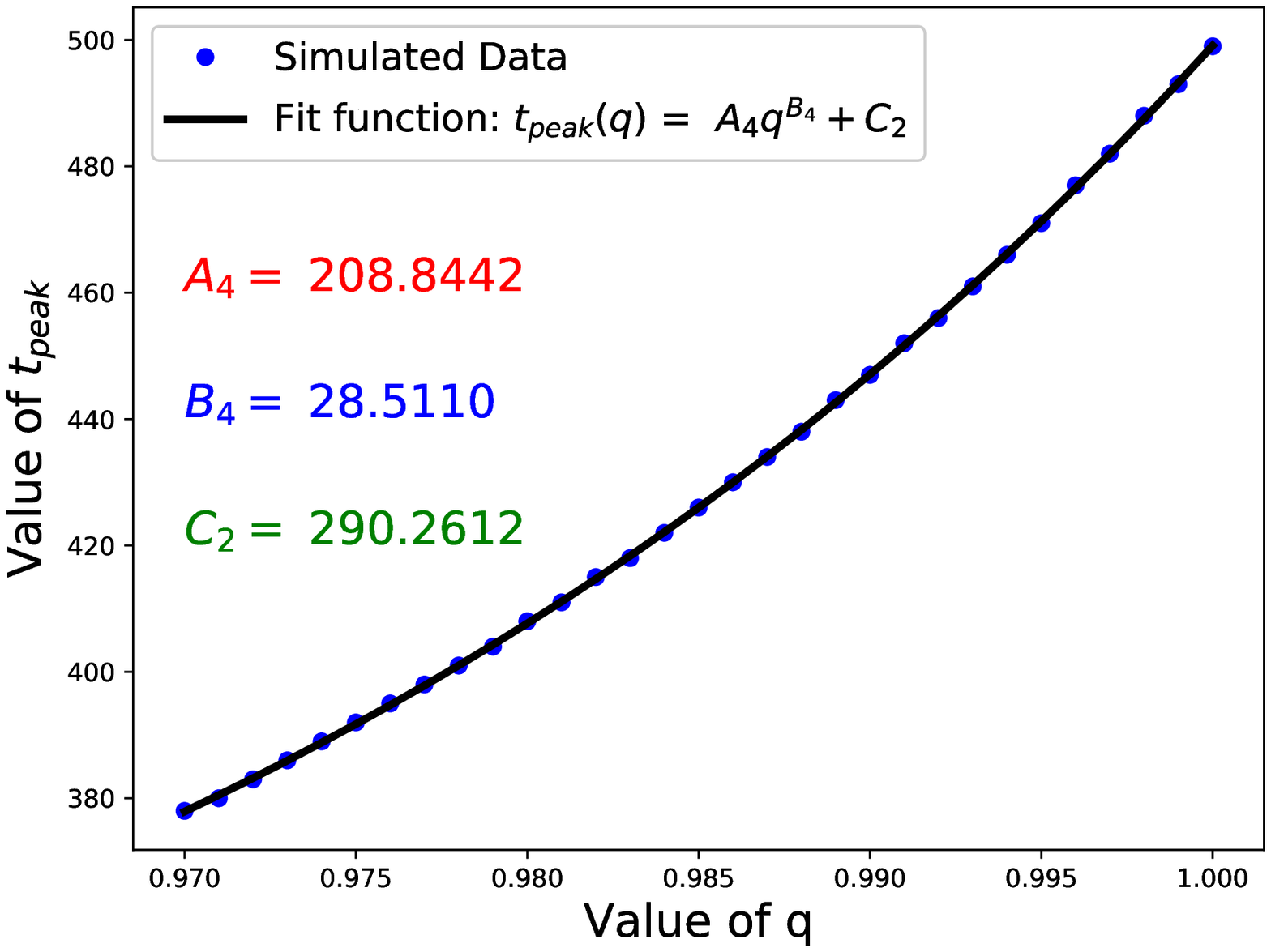}
 \caption{(Color online) (a) Fit of FIG.~\ref{fig:q=1}(c) in suitable range (b) Fit of FIG.~\ref{fig:q=1}(d) in suitable range (c) Fit of FIG.~\ref{fig:p=1}(c) in suitable range (d) Fit of FIG.~\ref{fig:p=1}(d) in suitable range.}
 \label{fig:fit}
\end{figure*}

From FIG.~\ref{fig:q=1} (and Simulation 2), it can be clearly observed that in Case-I, for $p >$ 0.972, we have non-zero $t_{peak}$, which reflects a situation of epidemic. But for $p <$ 0.958, we have $t_{peak}$ = 0 which reflects a situation of a non-epidemic disease, in spite of having $R_{0} >$ 1. But even if we consider only the epidemic case, the variation of $t_{peak}$ in sub-linear and super-linear regime occurs differently. When $p$ lies between 0.972 and 1, we see a linear fall in $t_{peak}$ with increase in $p$. But in the super-linear regime, when $p >$ 1, we see a more rapid fall. Hence, we fit both the super-linear(for $p >$ 0.972) and sub-linear regions nonlinearly and linearly, respectively using the fit functions, 
\begin{align}
t_{peak}(p) &= A_{1}p^{-B_{1}} + C_{1}   \nonumber  \\
\nonumber  \\ 
t_{peak}(p) &= A_{2}p + B_{2}  \label{eqn:fitp}
\end{align}

as shown in FIG.~\ref{fig:fit}(a) and FIG.~\ref{fig:fit}(b).

The estimated parameters from the fit as stated in FIG.~\ref{fig:fit}(a) and FIG.~\ref{fig:fit}(b) tell us that $t_{peak}$ falls extremely rapid in the super-linear regime, resulting in a large number of infected patients at that time which will eventually lead to a lot of deaths due to limited medical infrastructure. So, a sub-linear regime is favourable for $p$. \par  

An interesting trend, however, occurs in the region where $p$ lies between 0.958 and 0.972. This essentially corresponds to a region of transit from non-epidemic to epidemic stage. \vspace{0.2in}

{\bf{Case-II: p = 1} \par \vspace{0.2in}}

Similarly from FIG.~\ref{fig:p=1} (and Simulation 3), we see that in Case-II, for $q <$ 1.026, we have non-zero $t_{peak}$, which reflects a situation of epidemic. But for $q >$ 1.042, we have $t_{peak}$ = 0 which gives rise to a situation of a non-epidemic disease, in spite of having $R_{0} >$ 1. And similarly just like the previous case, even in the epidemic zone, the variation of $t_{peak}$ in sub-linear and super-linear regime occurs differently. When $q$ lies between 0.97 and 1, we see a non-linear rise in $t_{peak}$ with increase in $q$. But in the super-linear regime, for $q <$ 1.026, we see a linear rise. Hence, we fit both the regions (for $q <$ 1.026) linearly and nonlinearly, respectively using the fit functions, 
\begin{align}
t_{peak}(q) &= A_{3}q + B_{3}  \nonumber \\  
\nonumber \\
t_{peak}(q) &= A_{4}q^{B_{4}} + C_{2}
\label{eqn:fitq}
\end{align}

as shown in FIG.~\ref{fig:fit}(c) and FIG.~\ref{fig:fit}(d).

So, with a similar argument as in the last section, we infer that a super-linear regime is favourable for $q$ from the parameters as stated in FIG.~\ref{fig:fit}(c) and FIG.~\ref{fig:fit}(d).  \par

A similar interesting region of transit from epidemic to non-epidemic occurs here as well, where $q$ lies between 1.026 and 1.042. \par

\begin{figure*}[htpb]
 \centering
 (a)\includegraphics[width=0.75\columnwidth]{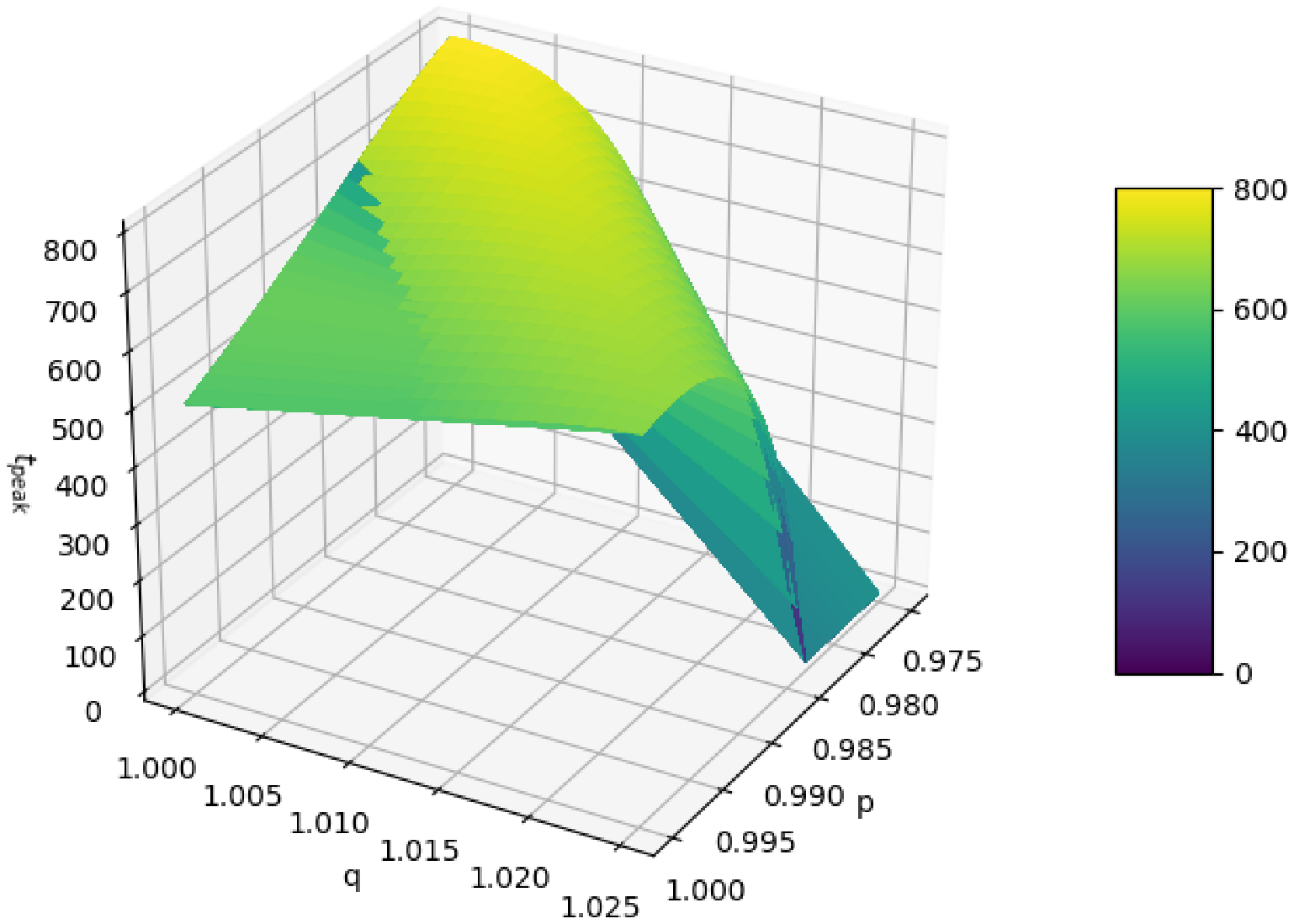}
 (b)\includegraphics[width=0.75\columnwidth]{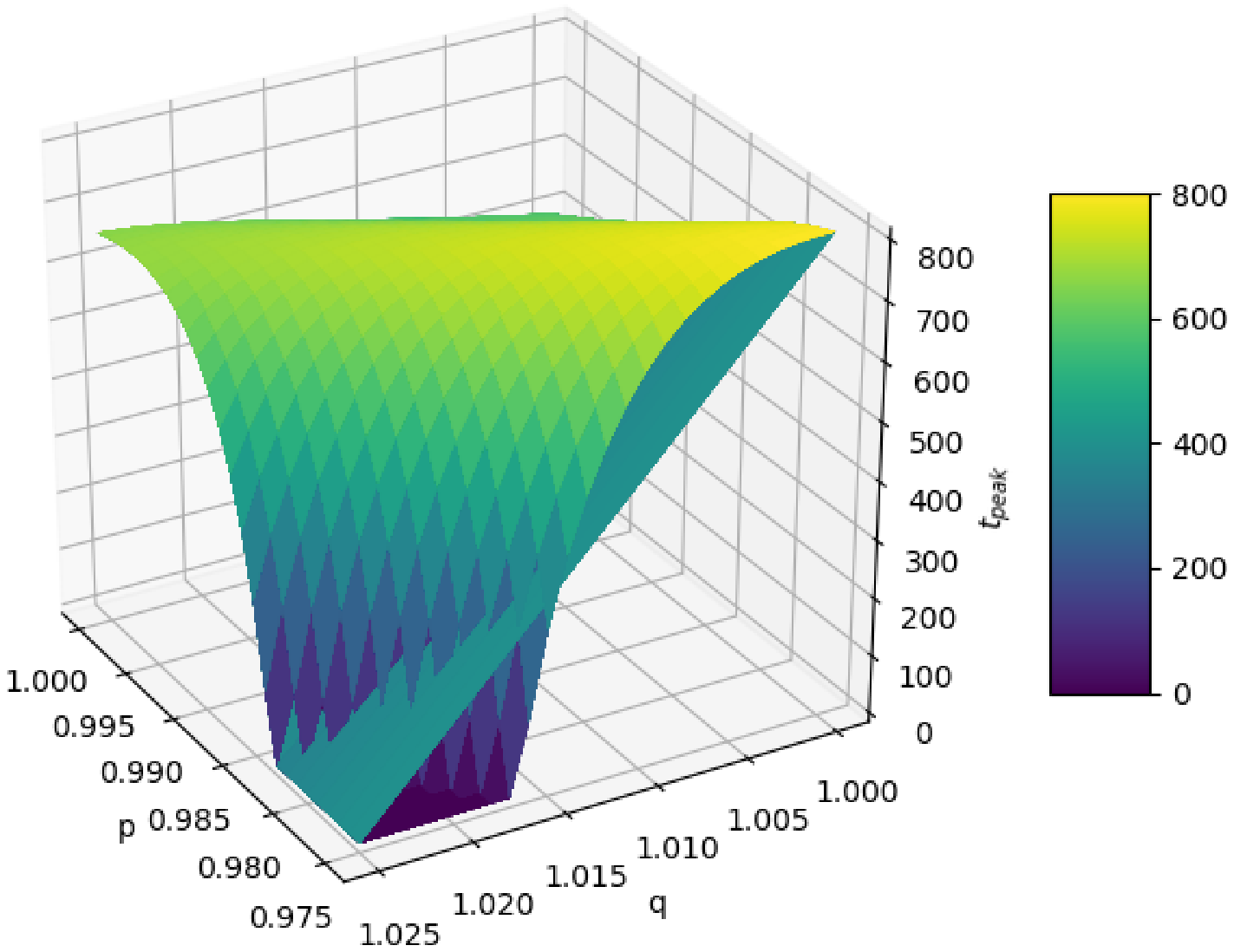}
 \\
 \caption{(Color online) 3D-surface plots of $t_{peak}$ as a function of $p$ and $q$ in regime of sub-linear $p$ and super-linear $q$ from two angles.} 
 \label{fig:3d}
\end{figure*}

Thus, we infer that a sub-linear perturbation of the infection exponent $p$ and a corresponding super-linear perturbation of the removal exponent $q$ will help us in battling an epidemic. We construct a 3-D surface plot for the aforesaid regions to visualize the situations properly, as shown in FIG.~\ref{fig:3d}.

A better visualization can be obtained by looking at the following simulation which shows the FIG.~\ref{fig:3d} in a coordinate system rotating with respect to the $t_{peak}$-axis.
Simulation 5:\textcolor{blue}{ \url{https://youtu.be/4KuuMLDmQxo}}

\subsubsection{Super-Linearity in q as a Problem of Phase Transition}
In statistical physics, we often explain the transformation of liquid-gaseous behaviour in materials, para-ferromagnetic transformation, etc. with the help of the theory of Phase Transition, where we look for a physical quantity that remains non-zero up to a certain critical point and beyond which it becomes zero. We refer to those physical quantities as "Order Parameter" $O_p$ . Now, in the super-linear regime of $q$, we see that below a critical value of $q$ ( $q$ = 1.042), $t_{peak}$ remains non-zero where there is the situation of epidemic and after $q$ = 1.042, $t_{peak}$ becomes zero, where there is no epidemic. So, by treating $t_{peak}$ as the order parameter, we can explain this phase transition from epidemic to non-epidemic state in the light of statistical mechanics. Thus, we obtain our critical value of $q$ to be $q_{c}$ = 1.042. Again, we argue that very close to $q_{c}$, $t_{peak}$ follows a power law, $(q_{c}-q)^{\beta}$, where $\beta$ is defined as the critical exponent. Now, in the following figure, FIG.~\ref{fig:pt}, we make a log-log plot of the variation of $t_{peak}$ as a function of $q$, very close to $q_{c}$ and fit it to a straight line using a least-square fitting to estimate the value of $\beta$.

From slope of the plot in FIG.~\ref{fig:pt}, we estimate the value of $\beta$ to be 0.654 $\pm$ 0.016. We propose, $O_p \sim (q_c-q)^{\beta}$, where $\beta=0.654 \pm 0.016$.

\subsection{Results from Incorporating Artificial Herd Immunity}
\begin{figure*}[htpb]
 (a)\includegraphics[width=0.46\columnwidth]{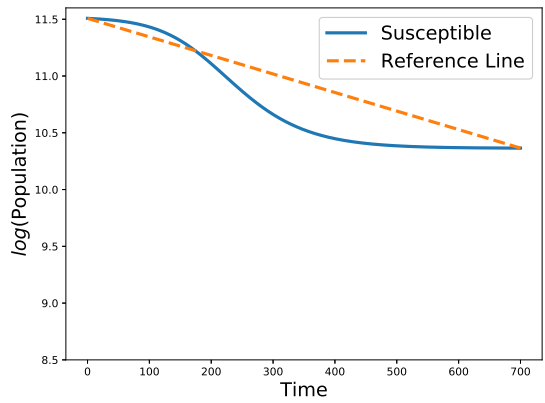}
 (b)\includegraphics[width=0.46\columnwidth]{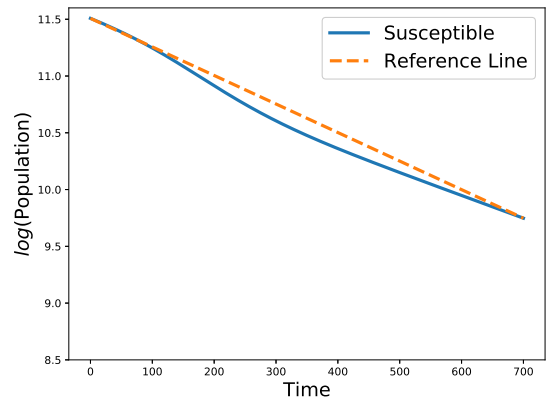} 
 \centering
 (c)\includegraphics[width=0.46\columnwidth]{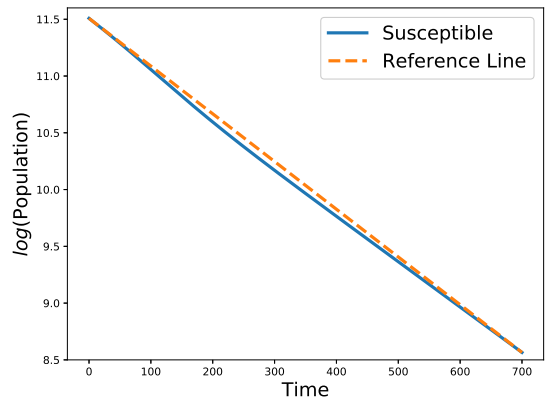}
 \caption{(Color online) Plots showing variation of logarithm of susceptible population as a function of time for three different values of $c$: (a) $c = 0$  (b) $c = 2 \times 10^{-3}$(c)  $c = 4 \times 10^{-3}$.}
 \label{fig:logherd}
\end{figure*}

\begin{figure}[h]
 \centering
 \includegraphics[width=0.88\columnwidth]{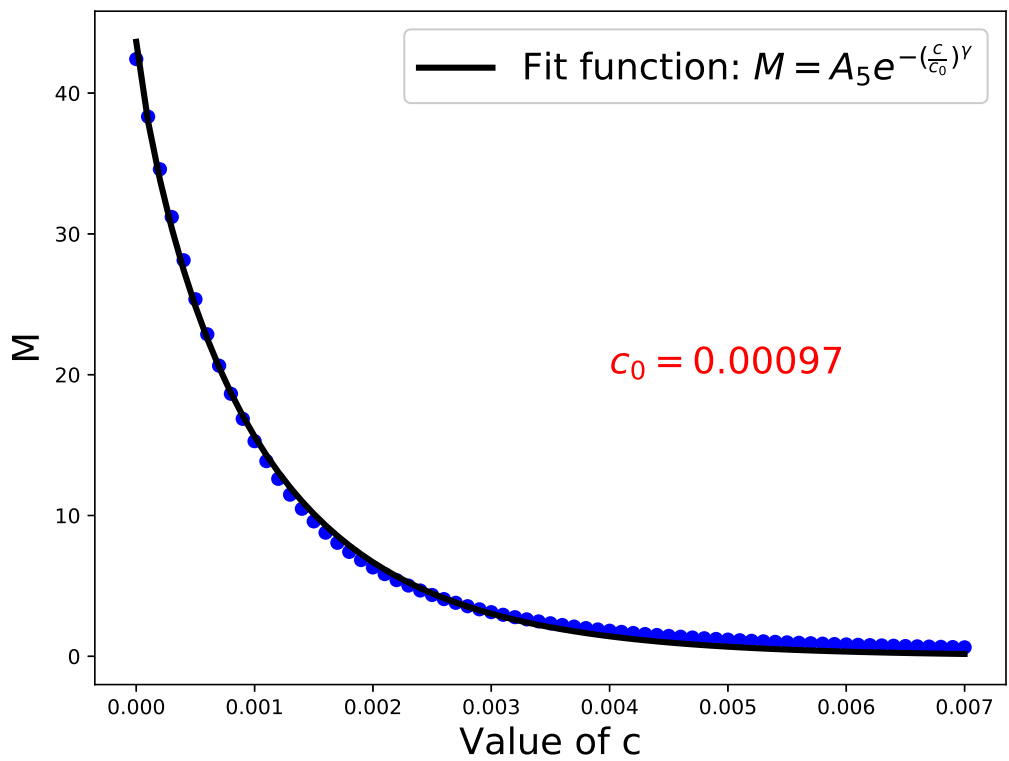}
 \caption{(Color online) Plot of mean square deviation of the logarithm of susceptible population from the reference line in FIG.~\ref{fig:logherd} and its chi-square fit to a stretched exponential.}
 \label{fig:msq}
\end{figure}

From FIG.~\ref{fig:herd} and Simulation 4, we observe that increase in the value of $c$ results in a change in the nature of time evolution of $x$ and $z$ accompanied by a decrease in both $t_{peak}$ and $y_{max}$, which is exactly similar to what we predict analytically. More vaccination will result in a faster decay of $x$, a faster growth in $z$ as well as a shorter span of the epidemic. Mathematically, from the plots, we see that the trend in the decay of $x$ with an increase in $c$ is indeed gradually becoming exponential. Now, we aim to estimate a scale value of the parameter $c$ for which we will get a reasonable prediction of the rate of vaccination, which needs to be done for ensuring a rapid eradication of the disease. In order to do that, we first plot the logarithm of the susceptible population with time along with a straight line corresponding to a situation of perfect exponential decay in FIG.~\ref{fig:logherd} for a series of $c$-values as shown in Simulation 6: \textcolor{blue}{ \url{https://youtu.be/Kvr_rB2o-hI}}.  Three frames of the simulation are shown in FIG.~\ref{fig:herd}.
We observe that the susceptible population indeed converges to the straight line as $c$ is increased just as we predicted. Then, we plot the mean square deviation of the logarithm of the population from the reference line, as shown in FIG.~\ref{fig:logherd} and fit the data points with a stretched exponential curve in FIG.~\ref{fig:msq} given by the following equation:

\begin{equation}
    M = A_{5}e^{-{(\frac{c}{c_{0}})}^{\gamma}}
    \label{eqn:fit2}
    \end{equation}

From FIG.~\ref{fig:msq} we estimate the values of Eqn.(\ref{eqn:fit2}) as $A_{5} = 43.64$, $\gamma = 0.87$. The critical value of $c$, i.e., $c_{0} = 0.001$. 
\vspace{0.3in}
\section{Conclusion} 

In summary, we did this work in two parts. In the first part, we have modified the SIR Model proposed by Kermack and McKendrick \cite{kermack} by incorporating two parameters, the infection exponent $p$ and the removal exponent $q$, given by Eqn.~\ref{eqn:MOD}. We have shown that small deviations of $p$ and $q$ about a linear trend, can give rise to a significant shift in the peak time of infection ($t_{peak}$). A sub-linear region of $p$ and a super-linear region of $q$ can not only pave the way for delaying the $t_{peak}$ so that we get enough time to discover vaccine but also can completely transform an infectious disease from an epidemic state to a non-epidemic state. We found this phenomenon of transformation analogous to the phase transition and estimate the suitable parameters associated with it, like the order parameter and the critical exponent. We found $O_p \sim(q_c-q)^{\beta}$, where $\beta=0.654 \pm 0.016$.
This is the key result of this article and has been observed and reported here first.

 In the second part, we have treated a situation where a vaccine is discovered and ready for use. We modified the model by considering an artificial herd immunity term which acts as a feedback loop in the time evolution of variables $x$ and $z$ such that, as the rate of vaccination $c$ increases, the susceptibility towards disease fall exponentially. We have solved the equations in Eqn.~\ref{eqn:vac} numerically and inferred that the susceptible population indeed tends to an exponential decay as the rate of vaccination is increased. In order to estimate a scale value of $c$, at first, we have plotted a logarithm of the susceptible population with respect to a reference line corresponding to a perfect exponential decay. Then we have plotted the mean square deviation of the logarithm of the susceptible population from the reference line as mentioned above and fitted it with a stretched exponential decay($M = A_{5}e^{-{(\frac{c}{c_{0}})}^{\gamma}}$). We have estimated the scale value of the rate of vaccination which has the potential to establish a sociological and clinical significance in the coming days, we believe.

\vskip 1cm
\textbf{Acknowledgement}
AD thanks Moumita Naskar for her help with the technicalities of the template of a research paper. AD also thanks Presidency University for allowing me to work in an academically engaging scientific environment. MA acknowledges FRPDF grant of Presidency University for financial support.


\newpage

\newpage

\newpage

\begin{figure}[h]
 \centering
 \includegraphics[width=0.88\columnwidth]{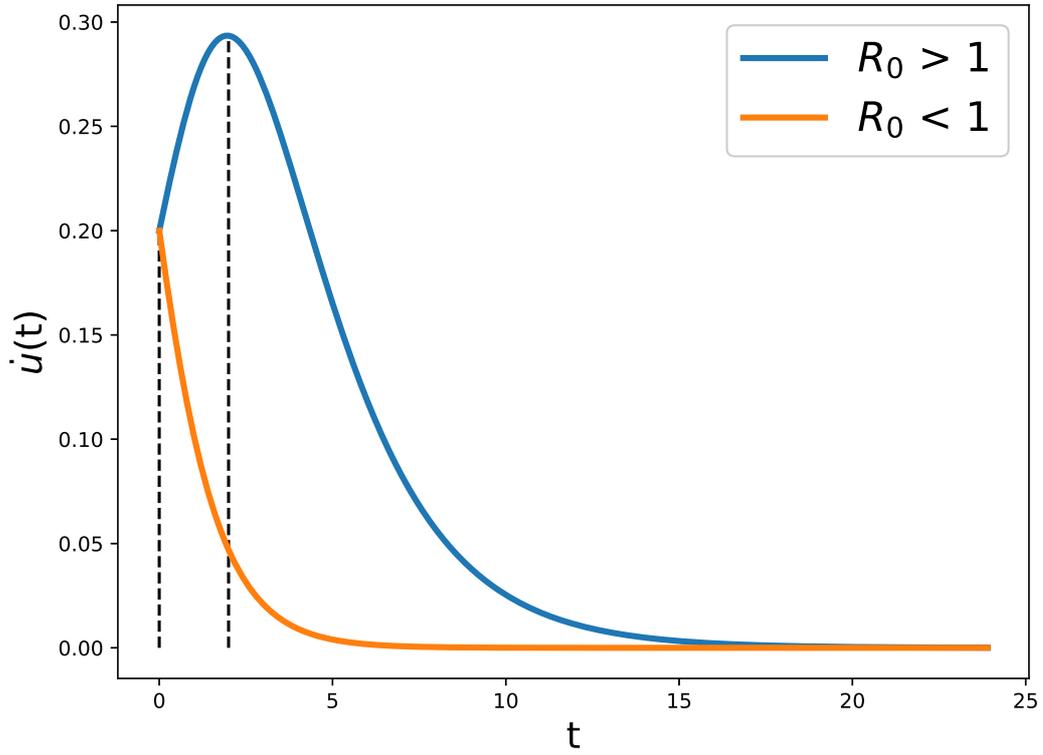}
 \caption{(Color online) When $R_{0} > 1$, we see, that $\dot{u}$ reaches a peak and then decreases. Since $y \propto \dot{z} \propto \dot{u}$, the value of t corresponding to this peak is nothing but the $t_{peak}$. So, we have a stage of epidemic. However, when $R_{0} < 1$, $\dot{u}$ and hence $y$ falls continually with time and gives us a condition of non-epidemic infectious disease.}
 \label{fig:r0}
\end{figure}

\newpage

\newpage

\newpage

\newpage

\newpage

\newpage

\newpage

\begin{figure}[h]
 \centering
 \includegraphics[width=0.88\columnwidth]{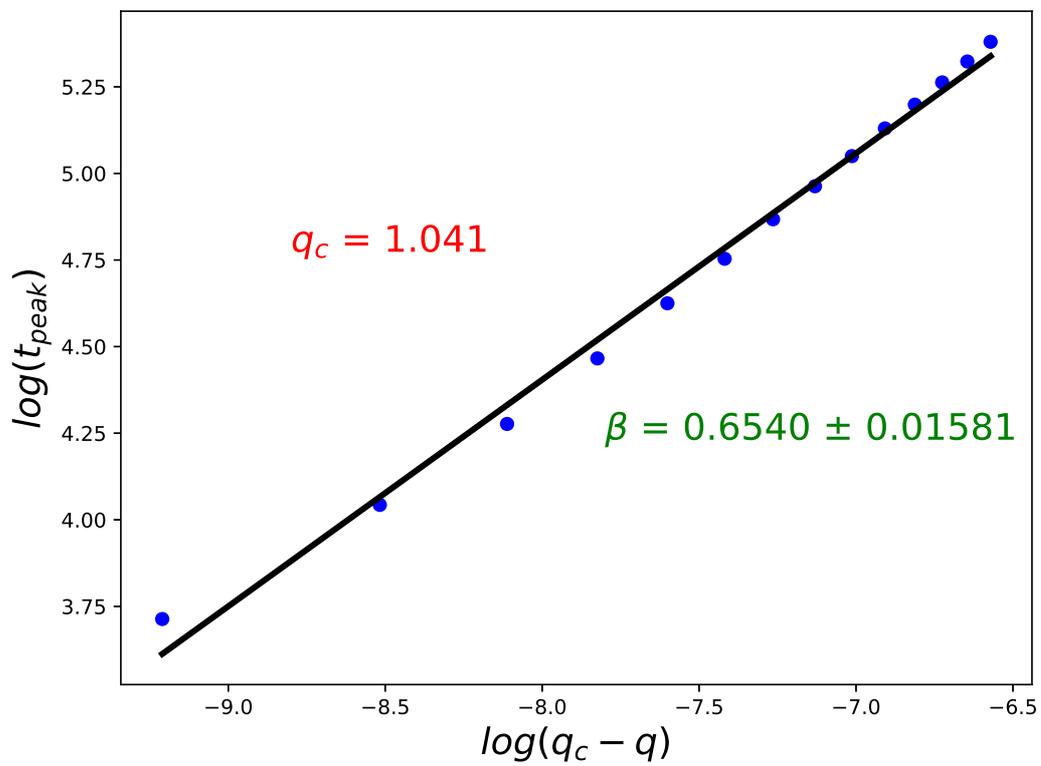}
 \caption{(Color online)Straight-line fit of the log-log plot of $t_{peak}$ vs $(q-q_{c})$ to estimate the value of $\beta$.}
 \label{fig:pt}
\end{figure}

\newpage

\newpage

\end{document}